\documentclass[11pt]{article}
\title{{Late time tails and nonlinear memories in asymptotically de Sitter spacetimes}}
\author{{\Large Yi-Zen Chu\textsuperscript{1,2}, M. Afif Ismail\textsuperscript{1}, and Yen-Wei Liu\textsuperscript{1,3}}\\ \\ {\Large \textsuperscript{1} Department of Physics, National Central University, Chungli 32001, Taiwan} \\ {\Large \textsuperscript{2} Center for High Energy and High Field Physics (CHiP),
	}\\ {\Large National Central University, Chungli 32001, Taiwan} \\ {\Large \textsuperscript{3} Department of Physics, National Tsing-Hua University, Hsinchu, Taiwan} } 
\date{}

\usepackage{graphicx}
\usepackage{multicol}
\usepackage{amsmath}
\usepackage{slashed}
\usepackage{young}
\usepackage{amsfonts}
\usepackage{amssymb}
\usepackage[usenames,dvipsnames]{color}
\usepackage{amsthm}
\usepackage{hyperref}
\usepackage[utf8]{inputenc} 
\usepackage[T1]{fontenc} 
\usepackage{XCharter} 
\usepackage{geometry} 

\theoremstyle{definition}

\numberwithin{equation}{subsection}

\newcommand{\GN}{G_{\rm N}}
\newcommand{\gb}{\overline{g}}

\newcommand{\dd}{\text{d}}

\definecolor{gray}{rgb}{0.4,0.4,0.4}
\definecolor{darkblue}{rgb}{0.0, 0.0, 0.55}

\newcommand{\vph}{\vphantom{\dot A}}

\begin{document}

\maketitle

\begin{abstract}
	We study the propagation of a massless scalar wave in de Sitter spacetime perturbed by an arbitrary central mass. By focusing on the late time limit, this probes the portion of the scalar signal traveling inside the null cone. 
	Unlike in asymptotically flat spacetimes, the amplitude of the scalar field detected by an observer at timelike infinity does not decay back to zero but develops a spacetime constant shift -- both at zeroth and first order in the central mass $M$. This indicates that massless scalar field propagation in asymptotically de Sitter spacetimes exhibits both linear and nonlinear tail-induced memories. On the other hand, for sufficiently late retarded times, the monopole portion of the scalar signal measured at null infinity is found to be amplified relative to its timelike infinity counterpart, by its nonlinear interactions with the gravitation field at first order in $M$.
\end{abstract}

\section{Introduction and Setup}

Non-linear perturbation theory about cosmological background is of physical interest. One of the reasons is, the asymptotic structure of cosmological spacetimes is quite different from that of flat background. In particular, linear perturbation theory on a de Sitter background is well studied \cite{deVega:1998ia} \cite{Date:2015kma}. For instance, Ashtekar, Bonga and Kesavan \cite{Ashtekar:2015lxa} found that adding a cosmological constant to the Einstein equation introduces new features to the quadrupole formula of gravitational waves. By integrating the equations-of-motion for the Weyl tensor, Bieri, Garfinkle and Yau \cite{Bieri:2015jwa} found the lightcone part of gravitational waves give rise to a memory effect similar to its Minkowski counterpart, except a redshift factor. Others \cite{Chu:2016ngc}, \cite{Tolish:2016ggo}, \cite{Kehagias:2016zry} have since confirmed and extended the results to more general Friedman-Robertson-Walker cosmologies. (Also, see \cite{Bonga:2020fhx}, \cite{Hamada:2017gdg}, \cite{Kehagias:2016zry}, \cite{Rojo:2020zlz} for studies of the asymptotic symmetries of cosmological spacetimes; and \cite{Chu:2019ssw} for memories in anti de Sitter spacetimes.) Being in de Sitter background, the gravitational waves do not just propagate strictly on the lightcone, but also inside the lightcone; this is known as the tail effect. One of us \cite{Chu:2015yua}, \cite{Chu:2016ngc} discovered that the gravitational wave tail also contributes to the memory effect. Now, about asymptotically flat spacetimes, it is known since Christodoulou \cite{Christodoulou:1991cr} and Blanchet and Damour \cite{Blanchet:1992br} that nonlinear corrections from General Relativity produces additional memories that at times might be even larger than the linear ones. In this paper, we initiate an examination of cosmological nonlinear memories by computing that of the massless scalar interacting gravitationally with an arbitrary central mass. Furthermore, we will assume the role of an observer approaching timelike infinity, to extract the tail-induced memories. As elucidated below, our analysis probes the scalar-graviton-scalar interaction, and is analogous to the de Sitter graviton $3$-point self-interactions that would therefore encode potential gravitational wave memories at leading nonlinear order. Finally, we will also compute the null infinity limit at late-retarded-times of the same setup.

The late time behavior of radiative fields in asymptotically-flat spacetimes has been well studied over the past decades. What is now known as ``Price's fall-off theorem" was first discovered by Price \cite{Price:1972pw}: the spin-0 radiative field decays in time according to an inverse power-law, with a power determined by the angular profile of the initial wave profile. In his work, Price is using the spherically symmetric Schwarzschild spacetime. The corresponding falloff properties of radiative fields in cosmological backgrounds is less well understood, although there are a number of numerical studies, such as Brady, Chambers, Krivan and Laguna (BCKL) \cite{Brady:1996za} and Brady, Chambers, Laarakkers and Poisson (BCLP) \cite{Brady:1999wd} on exact Schwarzschild-de Sitter and Reissner-Nordstr\"{o}m-de Sitter backgrounds.

In this paper, we want to investigate the simplest radiative field, a massless scalar field, in a perturbed de Sitter spacetime by using Poisson's approach to learn its late-time behavior \cite{Poisson:2002jz} analytically. Given an initial profile for the scalar wave packet, we found a constant tail signal observed at late times. This effect can be interpreted as tail induced memory effect to the scalar field. This complements and extends the results of BCKL \cite{Brady:1996za} and BCLP \cite{Brady:1999wd}. In \S \eqref{Section_Setup} we describe our setup; in \S \eqref{Section_Minkowski} we introduce a frequency space method to reproduce Poisson's work \cite{Poisson:2002jz} on the late time behavior of a scalar field in perturbed Minkowski background. Following which, in \S \eqref{Section_dS} we work on the scalar field solution in the late time regime of the perturbed de Sitter background.

\section{Setup}
\label{Section_Setup}

We consider a massless free scalar field $\Psi$, satisfying the equation
\begin{align}
	\begin{split}
		\Box_{x}\Psi [x] = 0, \ \ \ \ \ \ \ \,\ x^{\mu}=(x^{0},\vec{x}) ,
		\label{PsiWaveFunction}
	\end{split}					
\end{align}
where $\Box$ is the wave operator $\Box \equiv g^{\mu\nu}\nabla_{\mu}\nabla_{\nu}$. The geometry would take the following perturbed form\footnote{ Greek indices $\mu,\nu,....,$ run from $0$ (time) to $3$, while Latin ones $i, j, . . . ,$ run over only the spatial values $1$ to $3$ },
\begin{align}
	\label{PerturbedGeometry}
	g_{\mu\nu} &= a^2 \left( \eta_{\mu\nu} + \chi_{\mu\nu} \right) ,
	\qquad
	|\chi_{\mu\nu}| \ll 1 , \\
	\eta_{\mu\nu} &\equiv \text{diag}[1,-1,-1,-1] ;
\end{align}
where in the asymptotically Minkowski case
\begin{align}
	x'^0 = t' \in \mathbb{R}, \qquad x^0 = t \in \mathbb{R}, \qquad \text{and} \qquad a = 1;
\end{align}
whereas in the asymptotically de Sitter case
\begin{align}
	x'^0 = \eta', \qquad x^0 = \eta, \qquad \text{and} \qquad a[\eta] = -\frac{1}{H\eta} ,
\end{align}
with these cosmological conformal times running over the negative real line, $-\infty < \eta,\eta' < 0$, and $H>0$ is the Hubble constant.

\begin{figure}[h]
	\begin{center}
		\includegraphics[width=4.5in]{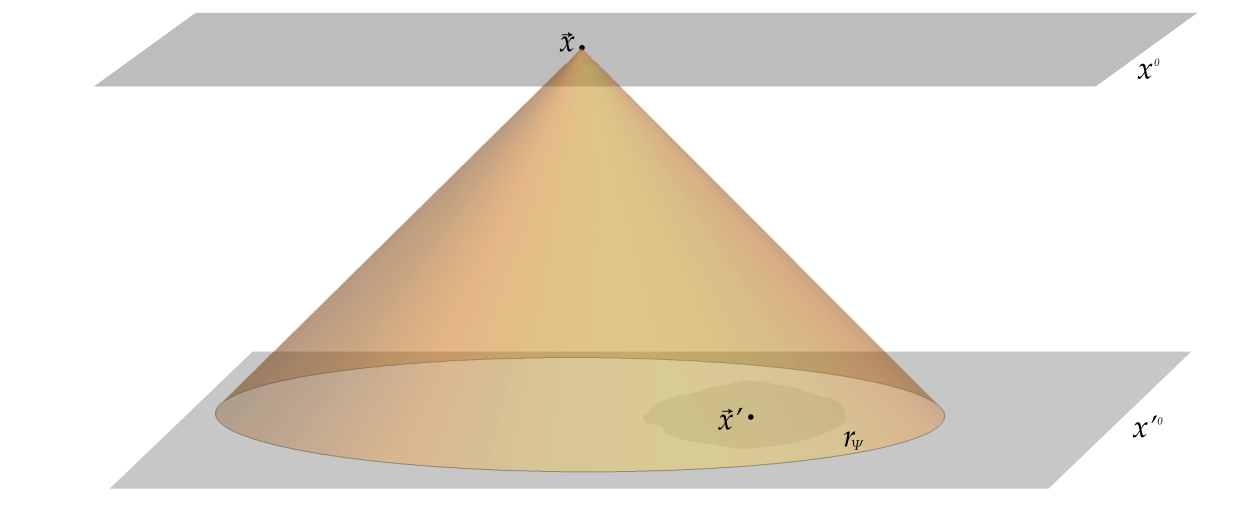}
		\caption{Localized initial scalar field profile and late time observer. Bottom and top planes are, respectively, the initial $x'^0$ hypersurface and final (observer) $x^0$ hypersurface. The observer is located at $\vec{x}$ and the cone denotes her past null cone. This illustrates, in the late time regime where the two hypersurfaces grow in distance, the observer continues to receive a scalar signal due solely to the tail effect.}
		\label{latetime}
	\end{center}
\end{figure}

Suppose at some initial time $x'^{0}$, the scalar field and its velocity are described by the following multipole expansion\footnote{We denote $\sum_{\ell,m} $ as the sum over {\it all} multipole indices, namely $\sum_{\ell=0}^{\infty} \sum_{m=-\ell}^{\ell} $, unless otherwise explained.},
\begin{align}
	\begin{split}
		\Psi[\vec{x}']=\sum_{\ell',m'} C_{\ell'}^{m'}[r'] Y_{\ell'}^{m'}[\widehat{x}'], \ \ \ \ \ \ \ \ \dot{\Psi}[\vec{x}']=\sum_{\ell',m'} \dot{C}_{\ell'}^{m'}[r'] Y_{\ell'}^{m'}[\widehat{x}'] ,
		\label{PsiInitialProfile}
	\end{split}					
\end{align}
where $Y_{\ell'}^{m'}[\widehat{x}']$ are spherical harmonics (which form basis orthonormal functions on the 2-sphere), and $C_{\ell'}^{m'}$ and $\dot{C}_{\ell'}^{m'}[r']$ are arbitrary functions that are non-zero only within some finite radii. This region where the initial conditions are non-zero is illustrated on the initial time hypersurface in Fig. \eqref{WavesCentralMassLightCones}. We will evolve this initial profile to the final time hypersurface $x^{0}$ by using Kirchhoff representation of the scalar field, up to first order in the perturbation $\chi_{\mu\nu}$:
\begin{align}
	\begin{split}
		\Psi[x^0 > x'^0,\vec{x}]
		&= \int_{\Bbb R^{3}} \dd^{3}\vec{x}' \sqrt{|h|} \widehat{n}^{\alpha'} \left( G[x,x'] \partial_{\alpha'} \Psi[x'] - \Psi[x']  \partial_{\alpha'}G[x,x'] \right) \\
		&= \int_{\Bbb R^{3}} \dd^{3}\vec{x}' a[\eta']^2 \left( 1 + \frac{1}{2} \eta^{ij} \chi_{ij} - \frac{1}{2} \chi_{00} \right) \left( G[x,x'] \dot{\Psi}[x'] - \Psi[x']  \partial_{0'} G[x,x'] \right) .
		\label{PsiKirchoff}
	\end{split}					
\end{align}
Here, $|h|$ is the determinant of induced metric on the constant time hypersurface, which has $\widehat{n}^{\alpha}$ as the normal vector.
\begin{align}
h_{\alpha\beta} \dd x^\alpha \dd x^\beta
&= -a^2 \left( \delta_{ij} - \chi_{ij} \right) \dd x^i \dd x^j  \\
\widehat{n}^\alpha &= a^{-1} \delta^\alpha_0 \left( 1 - \frac{1}{2} \chi_{00} + \mathcal{O}\left[\chi^2\right] \right)
\end{align}
The retarded Green function $G[x,x']$ describes the propagation of initial profile located at $x'$ to an observer at $x$ (see Fig. \eqref{latetime}) that obeys the wave equation
\begin{align}
	\begin{split}
		\Box G[x,x'] = \frac{\delta^{(4)}[x-x']}{\sqrt[4]{g[x]g[x']}} .
		\label{WaveEquationGreenFunction}
	\end{split}					
\end{align}
In \S \cite{Chu:2011ip}, a scheme was devised to solve Green's functions in a perturbed spacetime $g_{\mu\nu} = \gb_{\mu\nu} + h_{\mu\nu}$, in terms of the Green's functions in the unperturbed one $\gb_{\mu\nu}$. In our case, the massless scalar Green's function reads
\begin{align}
G[x,x'] = \overline{G}[x,x'] + \delta_1 G[x,x'] ;
\end{align}
where $\overline{G}$ solves the massless scalar wave equation with respect to $\gb_{\mu\nu} = a^2 \eta_{\mu\nu}$,
\begin{align}
\label{WaveEquationGreenFunction_Background}
\frac{\partial_\mu \left( a^2 \eta^{\mu\nu} \partial_\nu \overline{G} \right)}{a^4}
= \frac{\delta^{(4)}[x-x']}{a[x^0]^2 a[x'^0]^2} ;
\end{align}
and, in turn,
\begin{align}
\label{PerturbedGreensFunction}
\delta_1 G[x,x']
&= -\int \dd^4 x'' a[\eta'']^2 \partial_{\alpha''} \overline{G}[x,x''] \bar{\chi}^{\alpha''\beta''} \partial_{\beta''} \overline{G}[x'',x'] ;
\end{align}
with all double-primed indices denoting evaluation with respect to the integration variable $x''$; and the `trace-reversed' perturbation is defined as
\begin{align}
	\label{tracedreversedchi}
\bar{\chi}_{\mu\nu} \equiv \chi_{\mu\nu} - \frac{1}{2} \eta_{\mu\nu} \eta^{\alpha\beta} \chi_{\alpha\beta} ,
\end{align}
with its indices moved by the flat metric.

\begin{figure}[h]
	\begin{center}
		\label{FeynmanDiagram}
		\includegraphics[width=3in]{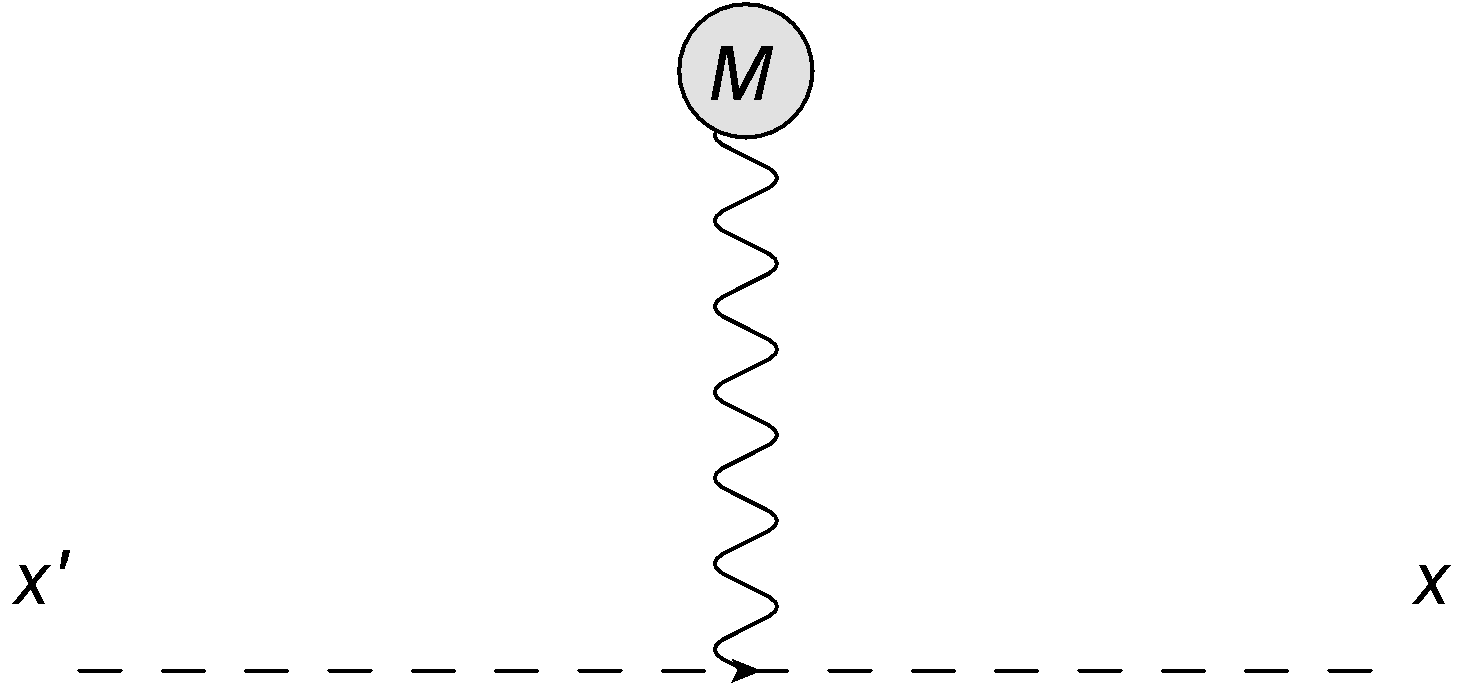}
		\caption{The Feynman diagram representing the massless scalar retarded Green's function at first order in the central mass. The dashed lines denote the zeroth order scalar Green's functions; whereas the wavy line tied to the blob labeled `$M$' describes the metric perturbation sourced by the static but otherwise arbitrary central mass distribution. From this depiction, we see that the scalar Green's function at first order in the central mass involves the scalar-graviton-scalar nonlinear interactions. (Drawn with JaxoDraw \cite{Binosi:2008ig}.)}
	\end{center}
\end{figure}

The metric perturbation $\chi_{\mu\nu}$ itself is sourced by the presence of an arbitrary quasi-static mass distribution consistent with energy-momentum conservation with respect to the background metric $\gb_{\mu\nu} = a^2 \eta_{\mu\nu}$, namely
\begin{align}
	\label{T00dS}
	T_{\mu\nu}[x''] = \frac{\rho[\vec{x}'']}{a[\eta'']} \delta_\mu^0 \delta_\nu^0 ;
\end{align}
This allows us to re-express the first order Green's function in eq. \eqref{PerturbedGreensFunction} as a Feynman diagram in Fig. \eqref{FeynmanDiagram} involving scalar-graviton-scalar interactions. In turn, this allows us to assert that the scalar memory revealed below is in fact of nonlinear character.

As an important aside: in a co-moving orthonormal frame, $T_{\widehat{0}\widehat{0}} = \rho/a^3$ and therefore the total mass with respect to the background metric is
\begin{align}
\label{TotalMass}
M
\equiv \int_{\mathbb{R}^3} T_{\widehat{0}\widehat{0}} a^3 \dd^3 \vec{x}'
= \int_{\mathbb{R}^3} \rho[\vec{x}''] \dd^3 \vec{x}'' .
\end{align}
All our major results below will be expressed in terms of this $M$.

We will focus on the late time limit of the scalar field in both Minkowski and de Sitter backgrounds. This means the difference between observation time $x^0$ and the initial time $x'^0$ obeys $x^0 - x'^0 \gg r + r'$, where $r \equiv |\vec{x}|$ and $r' \equiv |\vec{x}|$ are respectively the radial coordinates of the observer and an arbitrary point within the region where the initial scalar profile is non-trivial. Furthermore, in the Minkowski case, we will take both the null infinity limit, where the retarded time $u \equiv t-t'-r$ is held fixed and the advanced time $v \equiv t-t'+r$ is sent to infinity; as well as the timelike infinity limit, where the observer time is sent to infinity ($t \to \infty$) while her radial position $r$ is held still. In the de Sitter case, we will first consider the timelike infinity limit, where $\eta/\eta' \to 0$ while remaining well within the null cone, $\eta-\eta' \gg r > r'$. Furthermore, we shall assume that $-\eta'$ is the largest length scale in our problem: $-r'/\eta' < -r/\eta' \ll 1$. Then, we will examine the null infinity limit $v \equiv \eta-\eta'+r \to -2\eta'(1 + \mathcal{O}[u/\eta'])$ at fixed but late retarded times $u \equiv \eta-\eta'-r \gg r'$. As we shall witness in some detail below, that the observer continues to receive the scalar signal in such a late time regime, is due solely to the existence of tails -- see Fig. \eqref{latetime} -- i.e., a portion of the initial scalar field propagates inside the null cone.

\section{Perturbed Minkowski Background}
\label{Section_Minkowski}

As a warm up to de Sitter calculation, we first discuss the Minkowski background case. The late time behavior of a scalar field in a weakly curved spacetime was investigated by Poisson in \cite{Poisson:2002jz}. The calculation was performed by solving the Kirchhoff representation (\ref{PsiKirchoff}) directly in position space, and both the timelike and null infinity limits of the scalar solution were extracted.

Poisson extended Price's fall-off theorem of a scalar field by working in a perturbed Minkowski spacetime sourced by a central mass without any particular spatial symmetries, as opposed to the spherically symmetric black hole solution used by Price. In detail, Poisson assumed a spatially localized stationary mass distribution with mass density $T_{00} = \rho[\vec{x}]$ and mass-current density $T_{0i} = \vec j[\vec{x}]=\rho \vec v$, where $\vec v$ is the velocity within the matter distribution. By parametrizing eq. \eqref{PerturbedGeometry} as
\begin{align}
\dd s^2 = \left( 1 + 2 \Phi \right) \dd t^2
- \left( 1 - 2 \Phi \right) \dd\vec{x} \cdot \dd\vec{x}
- 8 ( \vec{A} \cdot \dd \vec{x} ) \dd t ,
\end{align}
the linearized Einstein's equations (with zero cosmological constant) reduce to a pair of Poisson's equations, whose solutions are
\begin{align}
\label{Minkowski_PoissonEqn_Phi}
\Phi[\vec{x}]
&= -16\pi \GN
\int_{\mathbb{R}^3} \frac{\rho[\vec{x}'']}{|\vec{x}-\vec{x}''|} \dd^3 \vec{x}'' \\
\vec{A}[\vec{x}]
&= -16\pi \GN
\int_{\mathbb{R}^3} \frac{\vec{j}[\vec{x}'']}{|\vec{x}-\vec{x}''|} \dd^3 \vec{x}''  .
\end{align}
He found the scalar field behavior at timelike infinity is given by
\begin{align}
	\begin{split}
		\Psi[t \gg r,r,\Omega]= 4G_\text NM(-1)^{\ell+1} \frac{(2\ell+2)!!}{(2\ell+1)!!} \frac{r^{\ell}}{t^{2\ell+3}} Y_{\ell}^{m}[\widehat{x}] \left( \dot{C}_{\ell} - \frac{2\ell+3}{t} C_{\ell} \right),
		\label{PoissonTimelike}
	\end{split}					
\end{align}
and at null infinity is given by
\begin{align}
	\begin{split}
		r\Psi[v \to \infty,u,\Omega] &= 2 G_\text N M (-1)^{\ell+1} \frac{(\ell+1)!}{(2\ell+1)!!} \frac{1}{u^{\ell+2}} Y_{\ell}^{m}[\widehat{x}] \left( \dot{C}_{\ell} - \frac{\ell+2}{u} C_{\ell} \right).
		\label{PoissonNull}
	\end{split}					
\end{align}
where $\widehat{x} \equiv \vec{x}/|\vec{x}|$ are the angular coordinates; and we use retarded time coordinate $u=t-r$ as well as advanced time coordinate $v=t+r$. (Here, Poisson has set $t' = 0$.) The $C_{\ell}$ and $\dot{C}_{\ell}$ are respectively defined to be $C_{\ell} \equiv \int \dd r\ r^{\ell+2} C_\ell^m[r']$ and $\dot{C}_{\ell} \equiv \int \dd r\ r^{\ell+2} \dot{C}_\ell^m[r']$.\footnote{Note that Poisson \cite{Poisson:2002jz} uses, respectively, $C$ and $\dot{C}$ in place of our $C_\ell^m$ and $\dot{C}_\ell^m$.}

The key finding of Poisson's results in equations \eqref{PoissonTimelike} and \eqref{PoissonNull} is that, even without assuming any spatial symmetries of the central mass, the falloff behavior of $\Psi$ is completely governed by its monopole moment in eq. \eqref{TotalMass} -- i.e., no higher mass multipoles entered the final answer -- and the power law in time depends on the multipole index of the initial data. We now proceed to reproduce the same results by employing a frequency space based method, which will also be employed in the de Sitter analysis in \S \eqref{Section_dS} below. The advantage of this approach is, it allows us to use a spherical harmonic basis from the outset.

\subsection{Frequency Space method}
\label{Section_MinkowskiFrequencySpace}

We will be content with reproducing Poisson's result involving the mass density only; for simplicity we will set $\vec{j}$ and hence $\vec{A}$ to zero. The first step is to perform a spherical harmonic decomposition to the solution to eq. \eqref{Minkowski_PoissonEqn_Phi} using the relation
\begin{align}
	\begin{split}
		\label{EuclideanGreenFunction}
		\frac{1}{4\pi |\vec{x}''-\vec{x}'''|} &= \frac{1}{r_{>}} \sum_{\ell''=0}^{\infty} \sum_{m''=-\ell''}^{\ell''}  \frac{Y_{\ell''}^{m''}[\widehat{x}''] \overline{Y}_{\ell''}^{m''}[\widehat{x}'''] }{2\ell''+1}  \left( \frac{r_{<}}{r_{>}}\right)^{\ell''},
	\end{split}					
\end{align}
where $r_{<}=\min[r'',r''']$ and $r_{>}=\max[r'',r''']$. This yields
\begin{align}
	\begin{split}
		\Phi[\vec{x}''] &= -16 \pi G_\text N \sum_{\ell'',m''} \frac{(-1)^{m''}}{2\ell''+1} \frac{Y_{\ell''}^{m''}[\widehat{x}'']}{r''^{\ell''+1}}M_{\ell''}^{m''},
		\label{PhiMultipoleExpand}
	\end{split}					
\end{align}
where
\begin{align}
	\begin{split}
		M_{\ell''}^{m''} \equiv \int_{0}^{r_{\ell''}^{m''}} \dd r''' r'''^{2+\ell''} \int_{\mathbb S^2} \dd \Omega''' Y_{\ell''}^{m''}[\widehat{x}'''] \rho[{\vec{x}'''}]
	\end{split}					
\end{align}
is the mass multipole moment. The mass density is localized inside radius $r_{\ell''}^{m''}$ of degree $\ell''$ and azimuthal index $m''$ and $\int \dd\Omega'''$ is the solid angle integral over the 2-sphere.

\textbf{Zeroth Order Green Function} \qquad In Minkowski spacetime, the propagation of a massless scalar field is described by the following retarded Green function
\begin{align}
	\begin{split}
		\overline{G}[x,x'] &=  \frac{\delta[t-t'-|\vec{x}-\vec{x}'|]}{4\pi |\vec{x}-\vec{x}'|},
\label{MasslessScalarGreensFunction_Minkowski}
	\end{split}					
\end{align}
that obeys the wave equation \eqref{WaveEquationGreenFunction_Background}, with $a=1$; namely, $\partial^2 \overline{G} = \delta^{(4)}[x-x']$. By re-expressing the delta function using its integral representation, we obtain the frequency space retarded Green function.
\begin{align}
	\overline{G}[x,x']
	&= \int_{-\infty}^{\infty} \frac{\dd\omega}{2\pi} e^{-i\omega(t-t')} \widetilde{G}^{+}[\omega;\vec{x}-\vec{x}']	,			
\end{align}
where
\begin{align}
	\begin{split}
		\widetilde{G}^{+}[\omega;\vec{x}-\vec{x}'] \equiv \frac{\exp[i\omega|\vec{x}-\vec{x}'|]}{4\pi |\vec{x}-\vec{x}'|} .
		\label{GFreqFlat}
	\end{split}					
\end{align}
The spherical harmonic decomposition of eq. \eqref{GFreqFlat} is
\begin{align}
	\begin{split}
		\widetilde{G}^{+}[\omega;\vec{x}-\vec{x}'] = i\omega \sum_{\ell,m} j_{\ell}[\omega r_{<}] h_{\ell}^{(1)}[\omega r_{>}] Y_{\ell}^{m}[\widehat{x}]  \overline{Y}_{\ell}^{m}[\widehat{x}'] ,
		\label{GFreqFlat1}
	\end{split}					
\end{align}
where $ j_{\ell}[\omega r_{<}]$ is the spherical Bessel function; $h_{\ell}^{(1)}[\omega r_{>}]$ is the Hankel function of the first kind; and the radii denote $r_{<}\equiv \min[r,r']$ and $r_{>}\equiv \max[r,r']$.

\textbf{First Order Green Function} \qquad We will now include the perturbation to the Green's function generated by the gravitational potential $\Phi$. This potential will scatter the null signals transmitted by the zeroth order Green's functions, so the perturbed signals can be seen as tail propagation from the perspective of the observer at $\vec{x}$. This is why the signal can be observed at late times, and is to be contrasted against the pure light-cone propagation of the signal in the unperturbed Minkowski background.

\begin{figure}[h]
	\begin{center}
		\includegraphics[width=4.5in]{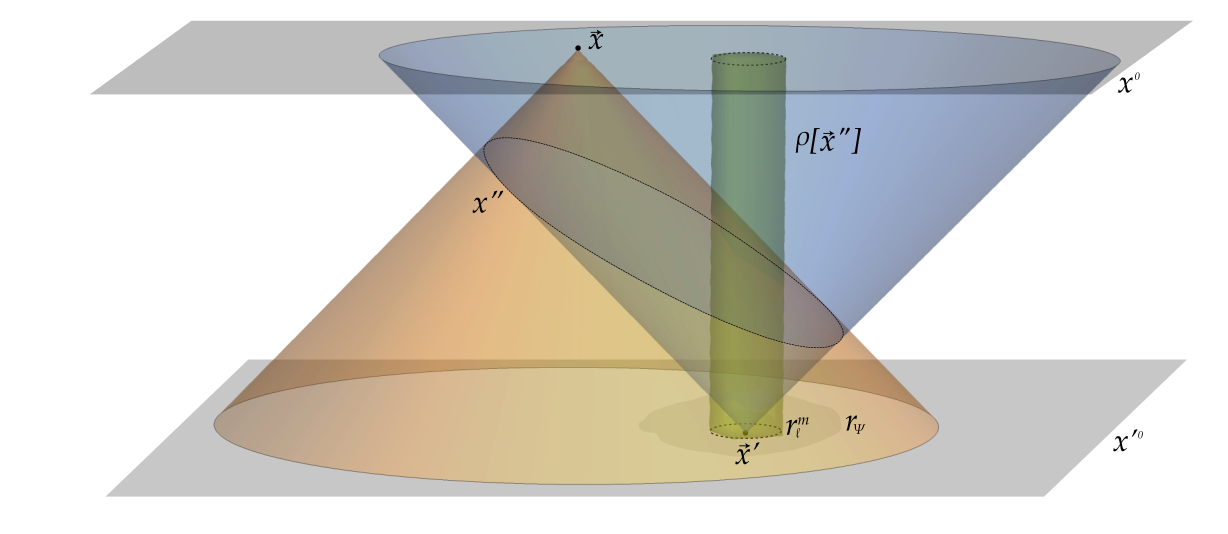}
		\caption{Spacetime diagram of the first order Green's function $\delta_1 G$. The bottom and top planes are, respectively, the initial and final time hypersurfaces. The cylindrical tube joining the two planes is the worldtube swept out by the central mass, whose proper mass density is $\rho[\vec{x}'']/a^3$, which we assume is zero outside some small radius. The shaded region on the initial time (lower) plane is where the initial $\Psi$ and its velocity are non-zero, which we too assume is a small region centered around the spatial origin. The observer on the final time (top) surface is located at $\vec{x}$, where we base her backward light cone. The first order massless scalar Green's function receives a signal emitted from the shaded region on the initial time surface, which propagates on the displayed forward null cone before scattering off the gravitational perturbation $\bar{\chi}_{00}[x'']$ engendered by the central mass, and taking another null path to the observer at $(\eta,\vec{x})$. The locus of spacetime points $\{x''\}$ where the scalar signal scatters off $\bar{\chi}_{00}$ is, by causality, given by the intersection of the forward light cone of $x' \equiv (x'^0,\vec{x}')$ and the backward light cone of $x \equiv (x^0,\vec{x})$. Notice, in the late time limit, the size of the ellipsoid will grow very large compared to $|\vec{x}'|$ and the spatial size of the mass distribution.}
		\label{WavesCentralMassLightCones}
	\end{center}
\end{figure}

The perturbed retarded Green function can be obtained from eq. \eqref{PerturbedGreensFunction} by setting $a=1$ and utilizing eq. \eqref{MasslessScalarGreensFunction_Minkowski}. Up to the first order in perturbation,
\begin{align}
	\begin{split}
		\label{delta1Gflat(00)}
		\delta_{1}G[x,x'] = - \int_{\mathbb R^{3,1}} \dd^{4}x'' \frac{\partial_{t''} \delta[t-t''-|\vec{x}-\vec{x}''|]}{4\pi |\vec{x}-\vec{x}''|} \Phi[\vec{x}'']  \frac{\partial_{t''} \delta[t''-t'-|\vec{x}''-\vec{x}'|]}{4\pi |\vec{x}''-\vec{x}'|}.
	\end{split}		
\end{align}
Evaluation of the time integral $t''$ over one of the Green functions will give us
\begin{align}
	\begin{split}
		\delta_{1}G[x,x'] = - \partial_{t}\partial_{t'} \int_{\Bbb R^{3}} \dd^{3}\vec{x}''  \frac{\delta[t-t'-|\vec{x}-\vec{x}''|-|\vec{x}''-\vec{x}'|]}{4\pi |\vec{x}-\vec{x}''| 4\pi |\vec{x}''-\vec{x}'|} \Phi[\vec{x}''] .
		\label{delta1Gflat(10)}
	\end{split}					
\end{align}
The delta function in eq.~\eqref{delta1Gflat(10)} indicates the null signals actually propagate from the initial point at $x'$, and reflects off the potential $\Phi$ at $\vec{x}''$ before reaching the observer at $x$. The volume integral $\vec{x}''$ must be evaluated on the intersection of future lightcone of the initial scalar field $x'$ and past lightcone of the observer at $x$, which is an ellipsoid with foci at $\vec{x}$ and $\vec{x}'$. By aligning the $\widehat{z}$ axis to be parallel to $\vec{x}-\vec{x}'$, the spatial components in a Cartesian basis of an arbitrary point of this ellipsoid reads \cite{Poisson:2002jz}, \cite{Chu:2011ip}
\begin{align}
\label{EllipsoidSurface}
\vec{x}''[\theta'',\phi''] &= \frac{\vec{x}+\vec{x}'}{2}
	+ \frac{1}{2} \sqrt{(t-t')^2-|\vec{x}-\vec{x}'|^2} \sin[\theta''] \widehat{e}_\perp[\phi'']
	+ \frac{t-t'}{2} \cos[\theta''] \widehat{z} .
\end{align}
The $\widehat{e}_\perp = (\cos\phi'',\sin\phi'',0)$ is the unit radial vector lying on the 2-dimensional plane orthogonal to $\vec{x}-\vec{x}'$. In Fig. \eqref{WavesCentralMassLightCones}, we illustrate with a spacetime diagram the causal structure of the scattering of null signals in eq. \eqref{delta1Gflat(10)}.

To decompose the integrand of (\ref{delta1Gflat(10)}) in terms of spherical harmonics, we again replace the delta function with its integral representation, with $\omega$ as the frequency.
\begin{align}
	\begin{split}
		\delta_{1}G[x,x'] = -  \int_{\Bbb R^{3}} \dd^{3}\vec{x}'' \int_{-\infty}^{\infty}\dd\omega \ \omega^{2}\ e^{-i\omega(t-t')} \frac{e^{i\omega|\vec{x}-\vec{x}''|}}{4\pi |\vec{x}-\vec{x}''|} \frac{e^{i\omega|\vec{x}''-\vec{x}'|}}{4\pi |\vec{x}''-\vec{x}'|} \Phi[\vec{x}''] ,
		\label{delta1Gflat(11)}
	\end{split}					
\end{align}
where $\omega^{2}$ is due to the $\partial_t \partial_{t'}$ in eq. \eqref{delta1Gflat(10)}. Since the expression (\ref{delta1Gflat(11)}) is time translation invariant, described by $\exp[-i\omega(t-t')]$, we can set the initial time $t'$ of the scalar field to be $0$. Employing the spherical harmonic decompositions of equations (\ref{GFreqFlat1}) and (\ref{PhiMultipoleExpand}), we may re-cast eq.~\eqref{delta1Gflat(11)} into
\begin{align}
	\delta_{1}G[x,x'] &=  \int_{\Bbb R^{3}} \dd^{3}\vec{x}'' \int_{-\infty}^{\infty}\dd\omega \ \omega^{2}\ e^{-i\omega t} 16 \pi G_\text N \sum_{\ell'',m''} \frac{(-1)^{m''}}{2\ell''+1} \frac{Y_{\ell''}^{m''}[\widehat{x}'']}{r''^{\ell''+1}}M_{\ell''}^{m''} \notag\\
	&\qquad\qquad \times i\omega \sum_{\ell,m} j_{\ell}[\omega r_{<}] h_{\ell}^{(1)}[\omega r_{>}] Y_{\ell}^{m}[\widehat{x}]  \overline{Y}_{\ell}^{m}[\widehat{x}''] \nonumber\\
	&\qquad\qquad \times i\omega \sum_{\ell',m'} j_{\ell'}[\omega r_{<}'] h_{\ell'}^{(1)}[\omega r_{>}'] Y_{\ell'}^{m'}[\widehat{x}'']  \overline{Y}_{\ell'}^{m'}[\widehat{x}'] \notag ,
\end{align}
where $r_{<} = \min[r,r'']$, $r_{>} = \max[ r,r'' ]$, $r_{<}' = \min[r',r'']$, and $r_{>}' = \max[r',r'']$. In the late time limit $u \equiv \Delta\eta-r \gg r'$, $r'$ will always be evaluated inside the boundary of the initial scalar field $r_{\Psi}$ (cf. Fig. \eqref{latetime}), which in turn is very small compared to $r''$, the spatial distance from the origin to $\vec{x}''$ on the ellipsoid. This is because, the ellipsoid size itself is controlled by $\Delta\eta \geq |\vec{x}-\vec{x}'| \approx r(1 + \mathcal{O}[r'/r])$. Therefore, we can directly set $r_{<}' = r'$ and $r_{>}' = r''$. For $r_{<}$ and $r_{>}$, will be determined by further calculation. Since $r'$ is the smallest length scales in our problem, we expect the final answer to admit a power series in $r'$. Motivated by this observation, we take the small argument limit of the spherical Bessel functions $j_{\ell'}[\omega r']$ and use integral representation of $j_{\ell}[\omega r]$,
\begin{align}
	\delta_{1}G[x,x'] &= -\int_{0}^{\infty} \dd r'' r''^{2} \int_{\mathbb S^2} \dd \Omega \int_{-\infty}^{\infty}\dd \omega\ \omega^{4} e^{-i\omega t}  16 \pi G_\text N \sum_{\ell'',m''}\frac{(-)^{m''}}{2\ell''+1} \frac{Y_{\ell''}^{m''}[\widehat{x}'']}{r''^{\ell''+1}}M_{\ell''}^{m''} \\
	&\hphantom{aaaa}  \times  \sum_{\ell,m} \int^{1}_{-1} \frac{\dd c}{2} \frac{e^{i \omega r_{<} c}}{\ell!} \left( -\frac{\omega r_{<}}{2} \right)^{\ell} (c^{2}-1)^{\ell} (-i)^{\ell+1} \frac{e^{i \omega r_{>}}}{\omega r_{>}} \sum_{s=0}^{\ell} \frac{i^{s}}{s! (2 \omega r_{>})^{s}} \frac{(\ell+s)!}{(\ell-s)!}  Y_{\ell}^{m}[\widehat{x}] \bar{Y}_{\ell}^{m}[\widehat{x}''] \notag \\
	&\hphantom{aaaa}  \times  \sum_{\ell',m'} \frac{(\omega r')^{\ell'}}{(2\ell'+1)!!} \left(1 + \mathcal{O}[(\omega r')^{2}] \right) (-i)^{\ell'+1} \frac{e^{i \omega r''}}{\omega r''} \sum_{s'=0}^{\ell'} \frac{i^{s'}}{s'! (2 \omega r'')^{s'}} \frac{(\ell'+s')!}{(\ell'-s')!} Y_{\ell'}^{m'}[\widehat{x}''] \bar{Y}_{\ell'}^{m'}[\widehat{x}'] .\notag
\end{align}
By collecting $\omega$ in the integrand, we can see that $\omega$ has positive power $2+\ell+\ell'-s-s'$. Therefore, $\omega^{2+\ell+\ell'-s-s'}$ can simply be replaced with $(i \partial_t)^{2+\ell+\ell'-s-s'}$, and the integral over $\omega$ evaluated.
\begin{align}
	\label{delta1Gflat(19)}	
	\delta_{1}G[x,x'] &= -   \sum_{\ell,m} \left( -\frac{ 1}{2} \right)^{\ell}\int^{1}_{-1} \frac{ \dd c}{2} \frac{1}{\ell!}  (c^{2}-1)^{\ell} (-i)^{\ell+1}  \sum_{s=0}^{\ell} \frac{i^{s}}{s! (2)^{s}} \frac{(\ell+s)!}{(\ell-s)!}  Y_{\ell}^{m}[\widehat{x}] (-)^{m}  \\
	&\hphantom{aaaaa} \times 32 \pi^{2} G_\text N \sum_{\ell'',m''}\frac{(-)^{m''}}{2\ell''+1}M_{\ell''}^{m''}     \sum_{\ell',m'} \frac{( r')^{\ell'}}{(2\ell'+1)!!}  (-i)^{\ell'+1}  \sum_{s'=0}^{\ell'} \frac{i^{s'}}{s'! (2 )^{s'}} \frac{(\ell'+s')!}{(\ell'-s')!}   \bar{Y}_{\ell'}^{m'}[\widehat{x}'] \notag \\
	&\hphantom{aaaaa} \times \frac{\partial_{t}^{2+\ell+\ell'-s-s'}}{(-i)^{2+\ell+\ell'-s-s'}}  \left\{  \frac{\Theta[t-c\ r-2 r]r^{\ell}}{(t-c\ r)^{\ell''+s'+s+1}}    \frac{1}{2} + \Theta\left[r-\frac{t-r}{1+c}\right] \left(\frac{t-r}{1+c}\right)^{\ell-\ell''-s'}   \frac{1}{ r^{s+1}}   \frac{1}{1+c} \right\} \notag \\
	&\hphantom{aaaaa} \times \left(\frac{(2\ell+1)(2\ell'+1)(2\ell''+1)}{4\pi} \right)^{1/2} \begin{pmatrix}
		\ell & \ell' & \ell'' \\
		0 & 0 & 0
	\end{pmatrix} \begin{pmatrix}
		\ell & \ell' & \ell'' \\
		-m & m' & m''
	\end{pmatrix} \left(1 + \mathcal{O}[( r'/t)^{2}] \right), \notag
\end{align}
where we have evaluated the solid angle integral $\Omega''$ with three spherical harmonics in the integrand, which introduces the Wigner 3j-symbols, and the $r''$ integral collapses the delta functions into step functions.

We will consider two possible configurations of the observer. The first: the observer is located near timelike infinity, $t \to \infty$ with $r$ fixed. In this case, the largest quantity is the observer time $t$. The second: the observer is located at null-infinity, where the advanced time grows without bound, $v \to \infty$ but the retarded time $u$ is large but fixed.

\textbf{Timelike infinity} \qquad Consider the observer at timelike infinity. For $r''>r$, we have a step function with $-2-c+t/r$ as the argument. Since we consider the observer at timelike infinity $t \rightarrow \infty$, the step function always has a positive value argument, i.e $ \Theta[(t/r-c) \gg 1]=1$. However, for $r>r''$ as $t \rightarrow \infty$, the argument is always negative, $\Theta[(c-t/r) \rightarrow -\infty]=0$, since the maximum value of cosine is $1$. Take the leading contribution of this limit, $(-c \ r + t) \rightarrow t$, followed by evaluating the integral,
\begin{align}
	\delta_{1}G[x,x'] &=    \sum_{\ell,m} \left( -\frac{ 1}{2} \right)^{\ell}  \frac{1}{\ell!}  \frac{\sqrt{\pi} \Gamma[\ell+1] }{\Gamma[\ell+3/2]} (-i)^{\ell+1}  \sum_{s=0}^{\ell} \frac{i^{s}}{s! (2)^{s}} \frac{(\ell+s)!}{(\ell-s)!}  Y_{\ell}^{m}[\widehat{x}] (-)^{m} \notag \\
	&\hphantom{aaa} \times 8 \pi^{2} G_\text N \sum_{\ell'',m''}\frac{(-)^{m''}}{2\ell''+1}M_{\ell''}^{m''}      \sum_{\ell',m'} \frac{( r')^{\ell'}}{(2\ell'+1)!!}  (-i)^{\ell'+1}  \sum_{s'=0}^{\ell'} \frac{i^{s'}}{s'! (2 )^{s'}} \frac{(\ell'+s')!}{(\ell'-s')!}   \bar{Y}_{\ell'}^{m'}[\widehat{x}'] \notag \\
	&\hphantom{aaa} \times  \frac{\Gamma[3+\ell+\ell'+\ell'']t^{-3-\ell-\ell'-\ell''}}{\Gamma[\ell''+s+s'+1](i)^{\ell+\ell'-s-s'}}  r^{\ell} \sqrt{\frac{(2\ell+1)(2\ell'+1)(2\ell''+1)}{4\pi} } \begin{pmatrix}
		\ell & \ell' & \ell'' \\
		0 & 0 & 0
	\end{pmatrix} \begin{pmatrix}
		\ell & \ell' & \ell'' \\
		-m & m' & m''
	\end{pmatrix} \notag \\
	&\hphantom{aaa} \times \left(1 + \mathcal{O}\left[ (r'/t)^{2}, r/t \right] \right).		
\end{align}
Parity invariance -- see Appendix \eqref{Section_Multipoles} -- tells us that only $\ell''=0$ mode survives and the double summation over $s$ and $s'$ is $(-1)^{\ell}$. The usual rules of angular momentum addition (as contained within the Wigner 3j-symbols) then informs us the surviving terms are those where $\ell=\ell'$.
\begin{align}
	\begin{split}
		\delta_{1}G[x,x'] =  4 G_\text N M \sum_{\ell,m}  t^{-2 \ell - 3}
		(r r')^{\ell}
		Y_{\ell}^{m}[\widehat{x}] \bar{Y}_{\ell}^{m}[\widehat{x}'] \frac{ (2 \ell+2)!! }{(2 \ell + 1)!!} (-1)^{\ell+1} \left(1 + \mathcal{O}[( r'/t)^{2},r/t] \right).
		\label{delta1GFlatTimelike}
	\end{split}					
\end{align}
We are going to use this Green function to describe the behavior of the scalar field. Recalling the Kirchhoff representation in eq. \eqref{PsiKirchoff}, with the initial data in eq. \eqref{PsiInitialProfile}, we arrive at
\begin{align}
	\begin{split}
		\Psi[x] &=  4 G_\text N M   t^{-2 \ell - 3}
		r ^{\ell}
		Y_{\ell}^{m}[\widehat{x}] \frac{ (2 \ell+2)!! }{(2 \ell + 1)!!} (-1)^{\ell+1} \\
		&\qquad \times \int_{0}^{\infty} \dd r'\ r'^{\ell+2}\left(  \dot C_{\ell}^{m}[r']  - \frac{2\ell+3}{t} C_{\ell}^{m}[r'] \right) .
	\end{split}					
\end{align}
Notice that we used time translation invariance on the second term, which introduced a minus sign. This solution is equivalent to Poisson's result (\ref{PoissonTimelike}).

\textbf{Null infinity} \qquad Now, let's consider the scalar field at null infinity. To do that, we introduce the retarded time coordinate $u \equiv t-r$ and advanced time coordinate $v \equiv t+r$ in (\ref{delta1Gflat(19)}), and proceed to consider the limits: $v \to \infty$ and $u$ fixed.
\begin{align}
	\delta_{1}G[x,x'] &= -    \sum_{\ell,m} \left( -\frac{ 1}{2} \right)^{\ell}\int^{1}_{-1} \frac{\dd c}{2} \frac{1}{\ell!}  (c^{2}-1)^{\ell} (-i)^{\ell+1}  \sum_{s=0}^{\ell} \frac{i^{s}}{s! (2)^{s}} \frac{(\ell+s)!}{(\ell-s)!}  Y_{\ell}^{m}[\widehat{x}] (-)^{m}   \\
	&\hphantom{aaaaa} \times 32 \pi^{2} G_\text N \sum_{\ell'',m''}\frac{(-)^{m''}}{2\ell''+1}M_{\ell''}^{m''}   \sum_{\ell',m'} \frac{( r')^{\ell'}}{(2\ell'+1)!!} (-i)^{\ell'+1}  \sum_{s'=0}^{\ell'} \frac{i^{s'}}{s'! (2 )^{s'}} \frac{(\ell'+s')!}{(\ell'-s')!}   \bar{Y}_{\ell'}^{m'}[\widehat{x}'] \notag \\
	&\hphantom{aaaaa} \times \frac{\partial_{u}^{2+\ell+\ell'-s-s'}}{(-i)^{2+\ell+\ell'-s-s'}}\left( \Theta \left[\frac{2u}{v-u}-1-c \right] \frac{2^{-\ell-1}(v-u)^{\ell}}{(1-c)^{\ell''+s'+s+1}} \right. \notag \\
	& \hphantom{aaaaaaaaaaaaaaaaaaaaa} \left.  + \Theta \left[1+c+\frac{2u}{u-v} \right] \frac{u^{\ell-\ell''-s'} 2^{s+1} }{(v-u)^{s+1}(1+c)^{\ell-\ell''-s'+1} }    \right) \notag \\
	&\hphantom{aaaaa} \times \left(\frac{(2\ell+1)(2\ell'+1)(2\ell''+1)}{4\pi} \right)^{1/2} \begin{pmatrix}
		\ell & \ell' & \ell'' \\
		0 & 0 & 0
	\end{pmatrix} \begin{pmatrix}
		\ell & \ell' & \ell'' \\
		-m & m' & m''
	\end{pmatrix} \left(1 + \mathcal{O}\left[ \left(r'/v\right)^{2},u/v \right] \right)	. \notag
\end{align}

The step function in the first term indicates that the $c$ integral bounded within $1>c>-1+2u/(v-u)$. After expanding the lower bound of the integration of the first term with respect to $2u/v$, the $c$ integral of the zeroth order is in fact the representation of beta function, while the first order is suppressed by $u/v$ relative to the zeroth order. Moreover, we also had the same situation in the second term. The integral is bounded within $-1+2u/(v-u)>c>-1$. Since there is a small window of the integration, $u/v \ll 1$, we need to treat the integral perturbatively by expanding the upper bound with respect to $2u/v$. After carrying out the $u$-derivatives, we found that the zeroth order of the expansion is zero, while the first order term is subleading by $u/v$ with respect to the zeroth order of the first term. Furthermore, because the summations here are similar to the timelike infinity calculation above, we were able to deduce that only the mass monopole $\ell''=0$ contributes at leading order, just like the timelike infinity case. Angular momentum addition rules then imply $\ell'=\ell$. 
\begin{align}
	\begin{split}
		\delta_{1}G[x,x'] = \sum_{\ell,m} 2 G_\text N M (-1)^{\ell+1} \frac{r'^{\ell}}{r \ u^{\ell+2}}  \frac{   (\ell+1)!! }{ (2
			\ell+1)!!} Y_{\ell}^{m}[\widehat{x}] \bar{Y}_{\ell}^{m}[\widehat{x}']  \left(1 + \mathcal{O}\left[ \left(r'/t\right)^{2},u/v \right] \right) .
		\label{delta1GflatNull}
	\end{split}					
\end{align}
We now use the form of the Green's function in eq. \eqref{delta1GflatNull} in the Kirchhoff representation of \eqref{PsiKirchoff} with eq. \eqref{PsiInitialProfile} as the initial profile of the scalar field. Finally, we have
\begin{align}
	\begin{split}
		r\ \Psi[v \to \infty,u,\Omega] &=  2 G_\text N M   \frac{1}{u^{\ell+2}} \frac{   (\ell+1)!
		}{ (2 \ell+1)!!
		}
		Y_{\ell}^{m}[\widehat{x}] \int_{0}^{\infty} \dd r'\ r'^{\ell+2}\left(  \dot C_{\ell}^{m}[r']  - \frac{\ell+2}{u} C_{\ell}^{m}[r'] \right) .
	\end{split}					
\end{align}
We have again used time translation invariance on the second term, which introduced a minus sign. This solution is equivalent to Poisson's result (\ref{PoissonNull}).

\section{Perturbed de Sitter Background}
\label{Section_dS}

\subsection{Pseudo-Frequency Space Method}
\label{Section_dS_Frequency}

We now turn to the de Sitter case, where $a[\eta] = -1/(H \eta)$. Within the generalized de Donder gauge condition $\eta \partial^\mu\overline\chi_{\mu\nu}= 2 \overline\chi_{0\nu}-\delta_\nu^0 \eta^{\rho\sigma}\overline\chi_{\rho\sigma}$, the metric perturbation $\chi_{\mu\nu}$ in eq. \eqref{PerturbedGeometry} (or, equivalently, its trace-reversed cousin $\bar{\chi}_{\mu\nu}$ in eq. \eqref{tracedreversedchi}) satisfying the linearized Einstein's equations with a positive cosmological constant $\Lambda = 3 H^2$ had been solved analytically in \cite{Chu:2016qxp} for a general source $T_{\mu\nu}$; namely
\begin{align}
	\delta_1 G_{\mu\nu} - \Lambda \ a^2 \chi_{\mu\nu} = 8\pi \GN T_{\mu\nu} ,
\end{align}
where $\delta_1 G_{\mu\nu}$ is the Einstein tensor expanded about a de Sitter background and containing precisely one power of $\chi_{\mu\nu}$. For technical simplicity we are considering a stress tensor in eq. \eqref{T00dS} whose only non-zero component is its mass density. Hence, the only non-zero component is $\bar{\chi}_{00}$, which can be found from the so-called \textit{pseudo-trace mode} solution in \cite{Chu:2016qxp}.
\begin{align}
	\begin{split}
		\bar{\chi}_{00}[\eta'',\vec{x}''] &= - \frac{16 \pi G_\text N}{ a[\eta''] } \int_{\Bbb R^{3}} \dd^{3}\vec{x}''' \frac{\rho[\vec{x}''']}{4\pi |\vec{x}''-\vec{x}'''|}
		\label{ChiBar00dS}
	\end{split}					
\end{align}
In terms of $\bar{\chi}_{00}$, the metric perturbation components are
\begin{align}
	\begin{split}
		\chi_{00} = \frac{1}{2} \bar{\chi}_{00}, \qquad
		\chi_{ij} = \frac{1}{2} \delta_{ij} \bar{\chi}_{00},\qquad
		\chi_{i0}=0.
	\end{split}					
\end{align}
Eq. \eqref{ChiBar00dS} tells us, up to the scale factor, $\bar{\chi}_{00}$ is identical to the flat spacetime Newtonian gravitational potential, which in turn means the spherical harmonic decomposition in eq. \eqref{EuclideanGreenFunction} may continue to be exploited. Specifically, if the $\bar{\chi}_{00}$ is evaluated outside the matter distribution then eq. \eqref{ChiBar00dS} becomes
\begin{align}
	\begin{split}
		\bar{\chi}_{00}[\eta'',\vec{x}''] &= - \frac{16 \pi G_\text N}{ a[\eta''] } \sum_{\ell'',m''}\frac{(-)^{m''}}{2\ell''+1}  \frac{ M_{\ell''}^{m''}[r_{\ell''}^{m''}]}{r''^{\ell''+1}} Y_{\ell''}^{m''}[\widehat{x}'']
		\label{ChiBar00dSOutside}
	\end{split}					
\end{align}
where the mass multipoles are 
\begin{align}
	\begin{split}
		M_{\ell''}^{m''}[r_{\ell''}^{m''}] &\equiv \int_{0}^{r_{\ell''}^{m''}}\dd r''' \int_{\mathbb S^2} \dd\Omega''' {r'''}^{\ell''+2} \overline{Y}_{\ell''}^{m''}[\widehat{x}''']\rho[\vec{x}'''] .
	\end{split}					
\end{align}
On the other hand, if $\bar{\chi}_{00}$ is evaluated inside the matter distribution, the integration needs to be separated into two different regions. The first region lies between the origin and the location of metric perturbation $r''$, where $r_{>}=r''$. However, the second region is between $r''$ and the boundary of matter distribution $r_{\ell''}^{m''}$, with $r_{>}=r'''$. Then, (\ref{ChiBar00dS}) becomes
\begin{align}
	\begin{split}
		\bar{\chi}_{00}[\eta'',\vec{x}''] &= - \frac{16 \pi G_\text N}{a[\eta''] } \sum_{\ell'',m''}\frac{(-)^{m''}}{2\ell''+1} Y_{\ell''}^{m''}[\widehat{x}''] \left( \frac{ M_{\ell''}^{m''}[r'']}{r''^{\ell''+1}} + N_{\ell''}^{m''}[r''] r''^{\ell''}\right)
		\label{ChiBar00dSInside}
	\end{split}					
\end{align}
where the internal multipoles are now
\begin{align}
	\label{InternalMultipoles}
	\begin{split}
		M_{\ell''}^{m''}[r''] &\equiv \int_{0}^{r''}\dd r''' \int_{\mathbb S^2} \dd\Omega''' {r'''}^{\ell''+2} \overline{Y}_{\ell''}^{m''}[\widehat{x}''']\rho[\vec{x}'''],\\
		N_{\ell''}^{m''}[r''] &\equiv \int_{r''}^{r_{\ell''}^{m''}} \dd r''' \int_{\mathbb S^2} \dd\Omega''' {r'''}^{-\ell''+1} \overline{Y}_{\ell''}^{m''}[\widehat{x}''']\rho[\vec{x}'''] .
	\end{split}					
\end{align}
These $M_{\ell''}^{m''}$ and $N_{\ell''}^{m''}$ contribute, respectively, to the two integration regions $0<r'''<r''$ and $r''<r'''<r_{\ell''}^{m''}$. 

\textbf{Initial Value Formulation} \qquad Now, let us prepare the Kirchhoff representation of the scalar field in a perturbed de Sitter background. Here, we set the initial hypersurface at $\eta'$. Therefore, up to first order in $\chi_{\mu\nu}$, the square root of induced metric determinant and normal vector of the initial hypersurface are
\begin{align}
	\begin{split}
	\sqrt{|h|} &\approx a[\eta']^{3} \left( 1- \frac{3}{4}  \bar{\chi}_{00} \right), \\
	\widehat{n}^{\alpha} &\approx a[\eta']^{-1} \delta^\alpha_0 \left( 1 - \frac{1}{4} \bar{\chi}_{00} \right) .
	\end{split}					
\end{align}
Up to first order in perturbations, we have from (\ref{PsiKirchoff}),
\begin{align}
	\Psi[x] &= \int_{\Bbb R^{3}}  \dd^{3}\vec{x}' a[\eta']^{2}  \left( \overline{G}[x,x'] \dot{\Psi}[x'] - \Psi[x']  \partial_{\eta'} (\overline{G}[x,x'] ) \right) \notag \\
	& \hphantom{aa} + \int_{\Bbb R^{3}}  \dd^{3}\vec{x}' a[\eta']^{2} ( -  \bar{\chi}_{00})  \left( \overline{G}[x,x'] \dot{\Psi}[x'] - \Psi[x']  \partial_{\eta'} (\overline{G}[x,x']) \right) \notag \\
	&\hphantom{aa} + \int_{\Bbb R^{3}}  \dd^{3}\vec{x}' a[\eta']^{2}   \left( (\delta_{1}G_{x,x'}) \dot{\Psi}[x'] - \Psi[x']  \partial_{\eta'} ( \delta_{1}G_{x,x'}) \right)  , \notag \\
	&\equiv \Psi^{(0)}[\eta,\eta',\vec{x}] + \Psi^{(0,\chi)}[\eta,\eta',\vec{x}]  + \Psi^{(1)}[\eta,\eta',\vec{x}] .
	\label{psipert}			
\end{align}
The first, second and third terms after the first equality are defined, respectively, as $\Psi^{(0)}[\eta,\eta',\vec{x}]$, $\Psi^{(0,\chi)}[\eta,\eta',\vec{x}]$, and $	\Psi^{(1)}[\eta,\eta',\vec{x}]$. $\Psi^{(0)}$ and $\Psi^{(0,\chi)}$ will be discussed in the next section. Unlike in 4D Minkowski, where there is no linear tail effect, these two terms do contribute to the observed scalar field at late times because the background de Sitter Green's function (in eq. \eqref{GreenFunctiondS} below) contains a tail. Therefore, we need to include these terms and compare them with the nonlinear signal $\Psi^{(1)}[\eta,\eta',\vec{x}]$ associated with scalar-gravity scattering described by $\delta_1 G$.

\textbf{Perturbed Green's Function} \qquad
The retarded Green's function of a massless scalar field in de Sitter spacetime contains two terms. One of them, proportional to a delta function, describes the propagation of the field on the lightcone; while the other, proportional to a step function, describes the propagation of the field inside the future lightcone of the source at $x'$. It is given by
\begin{align}
	\begin{split}
		\overline{G}[x,x'] &= \frac{H^{2}}{4 \pi} \left( \frac{\delta[\eta-\eta' - |\vec{x}-\vec{x}'|]}{|\vec{x}-\vec{x}'|} \eta \eta' + \Theta[\eta-\eta' - |\vec{x}-\vec{x}'|]  \right).
		\label{GreenFunctiondS}
	\end{split}					
\end{align}
The first order perturbed Green's function is given by eq. \eqref{PerturbedGreensFunction}. Its late time limit reads
\begin{align}
	\label{delta1GdS}
	&\delta_{1}G[x,x'] = -16 \pi G_\text N  \int_{0}^{\infty} \dd r'' r''^{2} \int_{\mathbb S^2} \dd \Omega''  H^{3} \   \sum_{\ell'',m''}\frac{(-)^{m''}}{2\ell''+1}\frac{Y_{\ell''}^{m''}[\widehat{x}'']}{r''^{\ell''+1}}    M_{\ell''}^{m''}  \\
	&\times \left( \eta\eta'  A_{-1}   - \eta    \eta' \partial_{\eta'} A_{0} +  \eta  B_{-1,\vec{x}}  -  \eta \eta' \partial_{\eta}A_{0} +  \eta \eta' \partial_{\eta}  \partial_{\eta'} A_{1}  -   \eta \partial_{\eta}B_{0,\vec{x}}  -   \eta' B_{-1,\vec{x}'} +  \eta' \partial_{\eta'} B_{0,\vec{x}'}  - C_{-1} \right)  \nonumber			
\end{align}
where we have defined
{\allowdisplaybreaks\begin{align}
		\label{Aa}
		A_{a} &\equiv \int_{-\infty}^{0} \dd \eta'' \eta''^{a}  \frac{\delta[\eta-\eta''-|\vec{x}-\vec{x}''|]}{4 \pi |\vec{x}-\vec{x}''|} \frac{\delta[\eta''-\eta'-|\vec{x}''-\vec{x}'|]}{4 \pi |\vec{x}''-\vec{x}'|}, \\
		\label{Bb}
		B_{b,\vec{x}_{o}} &\equiv \int_{-\infty}^{0} \dd \eta'' \eta''^{b}  \frac{\delta[\eta-\eta''-|\vec{x}-\vec{x}''|]\delta[\eta''-\eta'-|\vec{x}''-\vec{x}'|]}{(4 \pi )^{2}|\vec{x}_{o}-\vec{x}''|} ,\qquad \vec{x}_{o} \in \{ \vec{x},\vec{x}' \}, \\
		\label{Cm1}
		C_{-1} &\equiv \int_{-\infty}^{0} \dd \eta'' \frac{1}{(4 \pi )^{2}\eta''}  \delta[\eta-\eta''-|\vec{x}-\vec{x}''|] \delta[\eta''-\eta'-|\vec{x}''-\vec{x}'|] .
\end{align}}
That only delta functions are present in equations \eqref{Aa}--\eqref{Cm1} is because the derivatives in eq. \eqref{PerturbedGreensFunction} have converted the $\Theta$'s of the tail term in eq. \eqref{GreenFunctiondS} into $\delta$'s. Moreover, these delta functions teach us that, as far as the causal structure of the signal is concerned, the de Sitter massless scalar Green's function at first order in the central mass is similar to its asymptotically flat cousin: the signal at the observer's location $x$ arises from the signal emitted at $x'$ scattering off the central mass' gravitational potential lying on the ellipsoid that is associated with the intersection of the future light cone of $x'$ and the past light cone of $x$ -- recall Fig. \eqref{WavesCentralMassLightCones}. This also explains why, we have employed in eq. \eqref{delta1GdS} the form of the perturbation in eq. \eqref{ChiBar00dSOutside}; for, the ellipsoid of integration at late times always lies outside the matter distribution itself.

\textbf{Timelike infinity} \qquad In this work, we will be extracting the leading order terms from expanding in powers of $\eta/\eta'$, $r/\eta'$ and $r'/\eta' $. In particular, we shall specialize to an observer approaching timelike infinity, where $\eta/\eta' \to 0$ and $\eta-\eta' \gg r > r'$. By using these assumptions, the first six terms from the left are subleading relative to the remaining three terms:
\begin{align}
	\begin{split}
		\delta_{1}G[x,x']
		= -16 \pi G_\text N H^{3} \int_{0}^{\infty} \dd r'' \int_{\mathbb S^2} \dd\Omega''    \     &\sum_{\ell'',m''}\frac{(-1)^{m''}}{2\ell''+1}\frac{Y_{\ell''m''}[\widehat{x}'']}{r''^{\ell''-1}}    M_{\ell''}^{m''} \\
		&\times  \left( -  \eta'   B_{-1,\vec{x}'} +  \eta' \partial_{\eta'}  B_{0,\vec{x}'} - C_{-1}  \right).
		\label{delta1GdS(46)}
	\end{split}					
\end{align}
The calculations of $B_{-1,\vec{x}'}, B_{0,\vec{x}'}$ and $C_{-1}$ have been performed in the appendix \eqref{Section_ABC}. Here we will discuss how each term contributes. By including the $r''$ integration from $\delta_{1}G$, $B_{-1,\vec{x}'}$, $B_{0,\vec{x}'}$ and $C_{-1}$ have the following forms
\begin{align}
	\int_{0}^{\infty}  \dd r''  \frac{B_{-1,\vec{x}'}}{r''^{\ell''-1}} &= -\sum_{\ell,m} \frac{(r)^{\ell}}{(2\ell+1)!!}    Y_{\ell}^{m}[\widehat{x}]  \bar{Y}_{\ell}^{m}[\widehat{x}''] \sum_{\ell',m'} \frac{( r')^{\ell'}}{(2\ell' + 1)!!}    Y_{\ell'}^{m'}[\widehat{x}'']  \bar{Y}_{\ell'}^{m'}[\widehat{x}']  \frac{(-1)^{\ell+\ell'}(\ell+\ell'+\ell'')!}{ (\eta - \eta')^{\ell+\ell'+\ell'' +1}} \notag  \\
	&\times \frac{\pi  2^{1- \ell''} \Gamma [\ell'']  }{\Gamma
		\left[\frac{-\ell -\ell'+\ell''}{2} \right] \Gamma \left[\frac{ \ell -\ell'+\ell'' + 1}{2}\right] \Gamma \left[\frac{ -\ell+\ell'+\ell''+1}{2}\right] \Gamma
		\left[\frac{\ell +\ell'+\ell''+2}{2} \right]}  \left( 1 + \mathcal{O}\left[\frac{\eta}{\eta'}\right] \right)	
	\label{Bm1integrated}	,	\\
	\int_{0}^{\infty}  \dd r''  \frac{B_{0,\vec{x}'}}{r''^{-\ell''+1}} &=  \Bigg(    \frac{  \ 1 }{8\pi} \sum_{\ell',m'}  \frac{( r')^{\ell'}}{(2\ell'+1)!!}   Y_{\ell'}^{m'}[\widehat{x}'']  \bar{Y}_{\ell'}^{m'}[\widehat{x}'] \frac{2^{\ell''}}{(-1)^{\ell'} }  \frac{(\eta -\eta')^{-\ell'-\ell''} }{(\ell'')_{-\ell'} } \notag  \\
	&\hphantom{aa} + \sum_{\ell=1,m} \sum_{\ell',m'}  \frac{ 2^{\ell''-1} r^{\ell } r'^{\ell'}
		(-1)^{\ell + \ell'} (\eta - \eta')^{-\ell''-\ell -\ell'} }{ (2 \ell'+1)!!  }   \frac{  \Gamma [\ell''+\ell +\ell']}{(2 \ell +1)!! }  	\label{B0xpIntegrated}	   \\
	&\hphantom{aaaa} \times \frac{\pi  2^{3-2 \ell''} \Gamma [\ell''-1] Y_{\ell}^{m}[\widehat{x}]  \bar{Y}_{\ell}^{m}[\widehat{x}''] Y_{\ell'}^{m'}[\widehat{x}'']  \bar{Y}_{\ell'}^{m'}[\widehat{x}'] }{\Gamma
		\left[\frac{-\ell -\ell'+\ell''}{2} \right] \Gamma \left[\frac{ \ell -\ell'+\ell''-1}{2}\right] \Gamma \left[\frac{ -\ell+\ell'+\ell''+1}{2}\right] \Gamma
		\left[\frac{\ell +\ell'+\ell''}{2} \right]} \Bigg)  \left( 1 + \mathcal{O}\left[\left(\frac{r}{\eta'}\right)^{2}\right] \right) , \notag\\
	\int_{0}^{\infty}  \dd r''\frac{C_{-1}}{r''^{-\ell''+1}} &=  \Bigg( -    \frac{  \ 1 }{8\pi} \sum_{\ell,m}  \frac{( r)^{\ell}}{(2\ell+1)!!}   Y_{\ell}^{m}[\widehat{x}'']  \bar{Y}_{\ell}^{m}[\widehat{x}'] \frac{2^{\ell''}}{(-1)^{\ell} }  \frac{(\eta -\eta')^{-\ell-\ell''} }{(\ell'')_{-\ell} }  \nonumber	  \\
	&\hphantom{aa} - \sum_{\ell'=1,m'} \sum_{\ell,m}  \frac{ 2^{\ell''-1} r^{\ell } r'^{\ell'}
		(-1)^{\ell + \ell'} (\eta - \eta')^{-\ell''-\ell -\ell'} }{ (2 \ell+1)!!  }   \frac{  \Gamma [\ell''+\ell +\ell']}{(2 \ell' +1)!! }  \label{Cm1Integrated}   \\
	&\hphantom{aaaa} \times \frac{\pi  2^{3-2 \ell''} \Gamma [\ell''-1] Y_{\ell}^{m}[\widehat{x}]  \bar{Y}_{\ell}^{m}[\widehat{x}''] Y_{\ell'}^{m'}[\widehat{x}'']  \bar{Y}_{\ell'}^{m'}[\widehat{x}'] }{\Gamma
		\left[\frac{-\ell -\ell'+\ell''}{2} \right] \Gamma \left[\frac{ \ell' -\ell+\ell''-1}{2}\right] \Gamma \left[\frac{ -\ell'+\ell+\ell''+1}{2}\right] \Gamma
		\left[\frac{\ell +\ell'+\ell''}{2} \right]} \Bigg) \left( 1 + \mathcal{O}\left[\left(\frac{r'}{\eta'}\right)^{2}\right] \right) . \nonumber	
\end{align}
In the first term of eq.~(\ref{B0xpIntegrated}), because the Pochhammer symbol $(\ell'')_{-\ell'}$ will blow up if $\ell'=\ell''>0$, only the $\ell''=\ell=0$ terms survive. This in turn renders the first term of eq.~(\ref{B0xpIntegrated}) independent of both $\eta$ and $\eta'$, and since $B_{0,\vec{x}'}$ shows up in eq. \eqref{delta1GdS(46)} only as a derivative with respect to $\eta'$, we may therefore drop the first term of eq. \eqref{B0xpIntegrated}. Similar arguments related to $(\ell'')_{-\ell}$ in the first term of eq. (\ref{Cm1Integrated}) tell us only its $\ell=0$ term remains.

\begin{figure}[h]
	\begin{center}
		\includegraphics[width=6.0in]{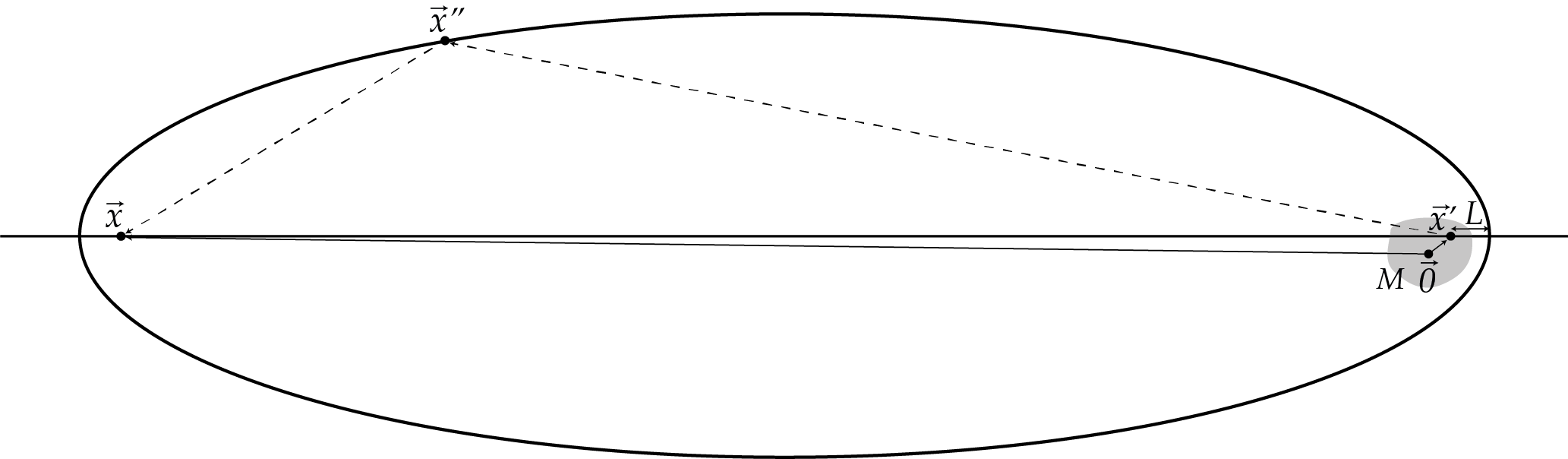}
		\caption{Spatial diagram of the first order Green's function $\delta_{1} G$. The grey blob denotes the central mass $M$. Referring to eq. \eqref{delta1Gflat(00)}, at first nonlinear order, the initial scalar field at $\vec{x}'$ propagates along null rays to $\vec{x}''$, scattering off the gravitational potential $\Phi[\vec{x}'']$ due to $M$, before taking null paths back to $\vec{x}$. The locus of $\vec{x}''$ is the surface of the ellipsoid given by causality: $\Delta \eta \equiv \eta-\eta' = |\vec{x}-\vec{x}'| + |\vec{x}' - \vec{x}''|$ -- arising from the argument of the $\delta$ function in eq. \eqref{delta1Gflat(10)}. The shortest distance between $\vec{x}''$ and $\vec{x}'$ is $L=(\Delta\eta-R)/2$. By expanding in powers of $r'$, we see that $L\approx (u/2)(1+\mathcal{O}[r'/u])$; i.e., the fractional change of $L$ is small if and only if $r' \ll u$.}
		\label{ellipsoid}
	\end{center}
\end{figure}

Additionally, the intermediate summations leading up to equations (\ref{Bm1integrated}), (\ref{B0xpIntegrated}) and (\ref{Cm1Integrated}) have been tackled in appendix \eqref{Section_DoubleSum} -- see the discussion leading up to eq. \eqref{SecondSum}. Now, the Gamma functions in the denominators of (\ref{Bm1integrated}), (\ref{B0xpIntegrated}) and (\ref{Cm1Integrated}) will blow up when their arguments are non-positive integers. 
In particular, parity arguments, spelt out in more detail in appendix \eqref{Section_ParityArgument}, provide the constraint $ \ell'' + 2 q = \ell + \ell'$, where $q$ is an arbitrary integer. This implies the $\Gamma\left[\frac{-\ell -\ell'+\ell''}{2} \right]$ in equations (\ref{Bm1integrated}), (\ref{B0xpIntegrated}) and (\ref{Cm1Integrated}) diverges, except when $\ell''=0$ due to the $\Gamma[\ell'']$ in eq. \eqref{Bm1integrated} and $\Gamma[\ell''-1]$ in equations \eqref{B0xpIntegrated} and \eqref{Cm1Integrated}. Angular momentum addition tells us $\ell = \ell'$ when $\ell''=0$; hence, at this point, we simply need to compute the limits
\begin{align}
	\lim_{\ell'' \to 0^+}  \frac{ 2^{1 - 2 \ell'' } \pi \Gamma[\ell''] }{\Gamma\left[ \frac{\ell''}2  -  \ell  \right] \Gamma\left[ \frac12 ( \ell''  + 1) \right] \Gamma\left[ \frac12 ( \ell'' + 1 )  \right] \Gamma\left[  \frac{\ell''}2 + (\ell + 1)  \right]  } = (-1)^\ell. \notag
\end{align}
and
	\begin{align}
	\lim_{\ell'' \to 0^+}  \frac{ 2^{2 - 2 \ell'' } \pi \Gamma[\ell''-1] }{\Gamma\left[ \frac{\ell''}2  -  \ell  \right] \Gamma\left[ \frac{ \ell''}{2} \right] \Gamma\left[ \frac12 ( \ell'' + 2 )  \right] \Gamma\left[  \frac{\ell''}2 + \ell  \right]  }
	& = 2\ell (-1)^\ell. \notag
\end{align}
Altogether, we gather
\begin{align}
	\delta_{1}G[x,x'] &=  \left( \delta_{1,0}G + \delta_{1,1}G[x,x'] \right) \left( 1 + \mathcal{O}\left[\frac{\eta}{\eta'} , \left(\frac{r}{\eta'}\right)^{2} , \left(\frac{r'}{\eta'}\right)^{2}\right] \right),				
	\label{delta1GdsFinal}
\end{align}
where we define the monopole-only term (with $\ell=\ell'=\ell''=0$) as
\begin{align}
	\label{delta10G}
	\delta_{1,0}G \equiv G_\text N  H^{3}    \frac{ M}{2\pi},
\end{align}
and the higher multipole terms as
\begin{align}
	\label{delta11G}
	\delta_{1,1}G[x,x'] = 2 G_\text N M  H^3   \sum_{\ell = 1}^\infty  &\sum_{m = -\ell}^\ell    (-1)^{\ell   }   \frac{ 2\ell!! }{ (2\ell + 1)!!  }  \left(\frac{ rr'}{ \eta'^2 }\right)^{\ell }    Y^{m}_\ell[\widehat{x}] \bar{Y}^{m}_\ell[\widehat{x}'].
\end{align}
\textbf{Null Infinity} \qquad We now turn to extracting the late retarded time and null infinity limits of $\delta_1 G$ in eq. \eqref{delta1GdS}. 
By late retarded time, we mean $u=\eta-\eta'-r \gg r'$ but is otherwise held fixed (see fig. \eqref{ellipsoid}); whereas the null infinity limit entails taking $\eta \to 0$ and the maximum possible advanced time coordinate allowed by the cosmological horizon, $v=\eta-\eta'+r \to -\eta'+r \approx -2\eta' (1+\mathcal{O}[u/\eta']) \gg u$.
Because $\eta \to 0$ in the null infinity limit, we find the first six terms to be subleading.
\begin{align}
		\delta_{1}G &= 16 \pi G  \int dr'' r''^{2} \int d\Omega''  \frac{1}{-H} H^{4} \  \left(   \sum_{\ell'',m''}\frac{(-)^{m''}}{2\ell''+1}\frac{Y_{\ell''m''}[\hat{x}'']}{r''^{\ell''+1}}    M_{\ell''}^{m''} \right) \nonumber\\
		&\hphantom{aaa} \times \left( \eta\eta'  A_{-1}   - \eta    \eta' \partial_{\eta'} A_{0} +  \eta  B_{-1,\vec{x}}  -  \eta \eta' \partial_{\eta}A_{0} +  \eta \eta' \partial_{\eta}  \partial_{\eta'} A_{1}  -   \eta \partial_{\eta}B_{0,\vec{x}} -   \eta' B_{-1,\vec{x}'} +  \eta' \partial_{\eta'} B_{0,\vec{x}'}  - C_{-1} \right)  ,  \nonumber\\
		\label{delta1GNullInfinity3Terms}
		&\approx 16 \pi G  \int dr'' r''^{2} \int d\Omega''  (-H^{3}) \  \left(   \sum_{\ell'',m''}\frac{(-)^{m''}}{2\ell''+1}\frac{Y_{\ell''m''}[\hat{x}'']}{r''^{\ell''+1}}    M_{\ell''}^{m''} \right) \nonumber \\
		&\qquad\qquad\qquad \times \bigg(-   \eta' B_{-1,\vec{x}'} +  \eta' \partial_{\eta'} B_{0,\vec{x}'}  - C_{-1} + \mathcal{O}[\eta/\eta'] \bigg) .
\end{align}
We shall focus only on the monopole contributions to the final signal.

To this end, we start by taking $B_{-1,\vec{x}'}$, $B_{0,\vec{x}'}$, and $C_{-1}$ from \eqref{Bm1a}, \eqref{B0a}, and \eqref{Cm1} respectively, followed by setting $\ell=\ell'=\ell''=0$.
\begin{align}
		-\eta'\int_{0}^{\infty}  \dd r''\ r''  B_{-1,\vec{x}'} &=
		\frac{v}{2} \int_{0}^{\infty}  \dd r''\ r'' \int_{-\infty}^{\infty} \frac{\dd\omega}{(4\pi)^{2}} \frac{\exp\left[-i\omega\left(\frac{u+v}{2}-r_{>}-r'' \right)\right] }{2\pi r_{>}r''} \bigg(\frac{e^{i\omega r_{<}}-e^{-i\omega r_{<}}}{2 i r_{<}\omega}\bigg) \notag \\
		&= -\frac{\ln[u/v]}{32\pi^{2}} \bigg(1+\mathcal{O}\bigg[\frac{u}{v}\bigg]\bigg)	,		
		\label{Bm1Null}	
\end{align}
\begin{align}
		\eta'\int_{0}^{\infty}  \dd r''\ r'' \partial_{\eta'} B_{0,\vec{x}'} &= -\frac{v}{2} \int_{0}^{\infty}  \dd r''\int_{-\infty}^{\infty} \frac{\dd\omega}{(4\pi)^{2}} \frac{i \exp\left[-i\omega(\frac{(u+v)}{2}-r_{>})\right]}{2\pi\omega r'' r} \notag \\
		& \qquad \times \left( -\frac{i}{2}\ \left( e^{-i \omega r_{<}}+e^{i \omega r_{<}} \right) r_{<}\omega + \frac{i}{2} \left(e^{-i \omega r_{<}}-e^{i \omega r_{<}}\right) (i+r_{<}\omega)  \right) \notag \\
		&= -\frac{v}{64 \pi ^2 u} + \frac{\ln[u/v]}{32 \pi ^2} +\mathcal{O}\bigg[\frac{u}{v}\bigg],
		\label{B0Null}
\end{align}
\begin{align}
		\int_{0}^{\infty}  \dd r''\ r''  C_{-1} &= 	\int_{0}^{\infty}  \dd r''\ r'' \int_{-\infty}^{\infty} \frac{\dd\omega}{(4\pi)^{2}} \frac{\exp\left[-i\omega\left( \frac{u+v}{2}-r_{>}-r'' \right)\right] }{2\pi r_{>}} \bigg(\frac{e^{i\omega r_{<}}-e^{-i\omega r_{<}}}{2 i r_{<}\omega}\bigg) \bigg( 1 + \frac{i r'}{3 r''^{2}\omega}\bigg) \notag \\
		&= \frac{1}{32\pi^{2}} \bigg(1+\mathcal{O}\bigg[\frac{u}{v}\bigg]\bigg)+\mathcal{O}\bigg[\frac{r'}{v}\bigg],
\end{align}
where $r_{<}=\min[r,r'']$ and $r_{>}=\max[r,r'']$. Inserting these integration results into eq. \eqref{delta1GNullInfinity3Terms}, we have
\begin{align}
\delta_{1}G &= \frac{G_{\text N}H^{3}M}{2\pi} \left( 1 + \frac{1}{2} \frac{v}{u} \right) .
\label{delta1GNull}
\end{align}
Observe that the first term in eq. \eqref{delta1GNull} is the same as its timelike infinity counterpart in eq. \eqref{delta10G}; whereas the second term in eq. \eqref{delta1GNull} describes a relative enhancement of $v/(2u)$.

\subsection{Position Spacetime Method}
\label{Section_dS_Position}
In the previous section, \S \eqref{Section_dS_Frequency}, we saw that at leading order in $\eta/\eta'$, $r/\eta'$, and $r'/\eta'$, the higher multipole moments of the central mass distribution did not contribute to the perturbed Green's function $\delta_1 G$. In this section, we will show how we may recover this mass-monopole-only result from a position spacetime calculation. In the late time limit, we have seen that $\bar{\chi}_{00}$ will never be evaluated inside the mass distribution, and hence the integrand of eq. \eqref{ChiBar00dS} may be Taylor expanded as follows
\begin{align}
\label{ChiBar00dS_Taylor}
\bar{\chi}_{00}[\eta'',\vec{x}'']
&= - \frac{4 G_\text N}{ a[\eta''] }\left( \frac{M}{|\vec{x}''|} + \sum_{\ell''=1}^{+\infty} \frac{(-)^{\ell''}}{\ell''!} \Xi_{00}^{\phantom{00}i_1 \dots i_\ell} \partial_{i''_1} \dots \partial_{i''_\ell} \frac{1}{|\vec{x}''|} \right)  , \\
\Xi_{00}^{\phantom{00}i_1 \dots i_\ell}
&\equiv
\int_{\Bbb R^{3}} \dd^{3}\vec{z}'' z''^{i_1} \dots z''^{i_{\ell}} \rho[\vec{z}''] .
\end{align}
This is equivalent to the spherical harmonic expansion of eq. \eqref{ChiBar00dS}, with the $\ell''$th term in the summation corresponding to the $\ell''$th mass multipole moment. Since we only wish to recover the mass monopole result of the previous section, we shall discard the summation in eq. \eqref{ChiBar00dS_Taylor} and insert it and eq. \eqref{GreenFunctiondS} into eq. \eqref{PerturbedGreensFunction}. By explicitly splitting the de Sitter Green's function $\overline{G}$ into the null cone $\overline{G}^{(\text{direct})}$ and tail $G^{(\text{tail})}$ pieces,
\begin{align}
	\begin{split}
	\overline{G}[x,x'] &= \overline{G}^{\text{(direct)}}[x,x'] + \overline{G}^{\text{(tail)}}[x,x'] , \\
	\overline{G}^{\text{(direct)}}[x,x'] &= \frac{H^{2}}{4 \pi} \frac{\delta[\eta-\eta' - |\vec{x}-\vec{x}'|]}{|\vec{x}-\vec{x}'|} \eta \eta' , \\
	\overline{G}^{\text{(tail)}}[x,x'] &= \frac{H^{2}}{4 \pi} \Theta[\eta-\eta' - |\vec{x}-\vec{x}'|] ;
	\end{split}					
\end{align}
this leads us to
\begin{align}
\label{PerturbedGreensFunction_dS_Position}
\delta_1 G[x,x'] = \sum_{1 \leq \text{I} \leq 4} \delta_1 G^{(\text{I})}[x,x'] ;
\end{align}
where
{\allowdisplaybreaks
	\begin{align}
		\hspace{-2em} \delta_1 G^{(1)}[x,x']  \equiv \, & - \eta^{\rho\mu''} \eta^{\sigma\nu''} \int_{-\infty}^0 \dd\eta'' \int_{\mathbb{R}^{3}}  \dd^3 \vec x'' \, a[\eta'']^{2} \, \partial_{\mu''} \overline G^{(\text{direct})}[x,x''] \, \overline \chi_{\rho\sigma}[ x'']\, \partial_{\nu''} \overline G^{(\text{direct})}[x'',x'] ,  \notag\\
		= \,& - \frac{ G_\text N M}{2\pi} H^3 \Bigg\{ \partial_\eta \partial_{\eta'} \left( \Theta[\eta - \eta'-R] \frac{\eta \eta'(\eta + \eta') }2 I_2  \right) \notag\\
		& - \partial_\eta \left( \Theta[\eta - \eta'-R]   \left( \eta\eta' I_1 + \frac{\eta (\eta + \eta')}2 I_2 \right) \right)  - \partial_{\eta'} \left( \Theta[\eta - \eta'-R]   \left( \eta\eta' I_1 + \frac{\eta'(\eta + \eta')}2 I_2 \right) \right)  \notag\\
		&+ \Theta[\eta - \eta'-R] \left( (\eta + \eta')I_1 + \frac{\eta + \eta'}2 I_2 + \frac{ 2\eta\eta'}{\eta + \eta'} I_3    \right) \Bigg\} , \\
		\delta_1 G^{(2)}[x,x']  \equiv \,&   - \eta^{\rho\mu''} \eta^{\sigma\nu''} \int_{-\infty}^0 \dd\eta'' \int_{\mathbb{R}^{3}}  \dd^3 \vec x'' \, a[\eta'']^{2} \, \partial_{\mu''} \overline G^{(\text{tail})}[x,x''] \, \overline \chi_{\rho\sigma}[ x'']\, \partial_{\nu''} \overline G^{(\text{direct})}[x'',x'] , \\
		= \,& - \frac{ G_\text N M}{ 4\pi}  H^3   \Bigg\{  \partial_{\eta'} \left(  \Theta[\eta - \eta' - R] \eta'  (\eta - \eta')  I_4^-  \vph \right)
		-  \Theta[\eta - \eta' - R]  (\eta - \eta')  \left(  \frac{2\eta'}{\eta + \eta'} I_5^- + I_4^- \right)   \Bigg\}  , \notag\\ 
		\delta_1 G^{(3)}[x,x']  \equiv \, & - \eta^{\rho\mu''} \eta^{\sigma\nu''} \int_{-\infty}^0 \dd\eta'' \int_{\mathbb{R}^{3}}  \dd^3 \vec x'' \, a[\eta'']^{2} \, \partial_{\mu''} \overline G^{(\text{direct})}[x,x''] \, \overline \chi_{\rho\sigma}[ x'']\, \partial_{\nu''} \overline G^{(\text{tail})}[x'',x'] , \\
		= \,& \frac{G_\text N M}{4\pi}  H^3 \Bigg\{  \partial_{\eta} \left( \Theta[\eta - \eta' - R] \eta (\eta - \eta') I^+_4 \vph \right)
		-  \Theta[\eta - \eta' - R] (\eta - \eta') \left( I_4^+ + \frac{2\eta}{\eta + \eta'} I^+_5 \right)   \Bigg\} , \notag \\
		\delta_1 G^{(4)}[x,x']  \equiv \, & - \eta^{\rho\mu''} \eta^{\sigma\nu''} \int_{-\infty}^0 \dd\eta'' \int_{\mathbb{R}^{3}}  \dd^3 \vec x'' \, a[\eta'']^{2} \, \partial_{\mu''} \overline G^{(\text{tail})}[x,x''] \, \overline \chi_{\rho\sigma}[ x'']\, \partial_{\nu''} \overline G^{(\text{tail})}[x'',x'] , \notag\\
		= \,&  \frac{G_\text N M}{4 \pi}  H^3 \left(  \Theta[\eta - \eta' - R]   \frac{ (\eta - \eta')^2 }{\eta + \eta'} I_6 \right) ,
\end{align}}%
and, by placing $\vec{x}''$ on the ellipsoidal surface parametrized in eq. \eqref{EllipsoidSurface} and defining $R \equiv |\vec{x}-\vec{x}'|$, the integrals involved are defined by
{\allowdisplaybreaks\begin{align}
I_1 &\equiv \int_{\mathbb{S}^2} \frac{\dd\Omega''}{4\pi} \frac{1}{|\vec{x}''|} ,  \label{I1toI6}\\
I_2 &\equiv \int_{\mathbb{S}^2} \frac{\dd\Omega''}{4\pi}  \,\frac{ 1 - \xi \cos\theta''}{|\vec{x}''|} , \qquad \xi \equiv - \frac R{\eta + \eta'}, \quad 0 < \xi <1 , \notag\\
I_3 &\equiv \int_{\mathbb{S}^2} \frac{\dd\Omega''}{4\pi}  \,\frac{1}{|\vec{x}''| (1 - \xi \cos\theta'')}  ,  \notag\\
I^\pm_4 &\equiv \int_{\mathbb{S}^2} \frac{\dd\Omega''}{4\pi}  \,\frac{1 \pm \zeta \cos\theta''}{|\vec{x}''|} , \qquad \zeta \equiv \frac{R}{\eta-\eta'}, \qquad 0 < \zeta < 1, \qquad \eta-\eta'>r+r',  \notag\\
I^\pm_5 &\equiv \int_{\mathbb{S}^2} \frac{\dd\Omega''}{4\pi}  \,\frac{1 \pm \zeta \cos\theta''}{|\vec{x}''| (1 - \xi \cos\theta'') }  ,  \notag\\
I_6 &\equiv \int_{\mathbb{S}^2} \frac{\dd\Omega''}{4\pi}  \,\frac{ ( 1 + \zeta \cos\theta'') (1 - \zeta \cos\theta'') }{|\vec{x}''| (1 - \xi \cos\theta'') } . \notag
\end{align}}
The $I_1$ was evaluated by DeWitt and DeWitt \cite{DeWitt:1964de}; for $\eta - \eta' > r+r'$, relevant for our late time calculations, it reads
\begin{align}
I_1 = \frac{1}{R} \ln \left[ \frac{1 + \zeta}{1 - \zeta} \right] .
\end{align}
{\bf Timelike infinity} \qquad In the timelike infinity limit, $-r/\eta',-r'/\eta',\eta/\eta' \ll 1$, both the $\xi$ and $\zeta$ tend to zero. As such, the rest of $I_{2,3,4,5,6}$ are really $I_1$ up to fractional corrections of order $-r/\eta'$.
\begin{align}
I_{2,3,4,5,6} = I_1 \left( 1 + \mathcal{O}[\zeta,\xi] \right).
\end{align}
Taking all these into account, and returning to our perturbed Green's function in eq. \eqref{PerturbedGreensFunction_dS_Position},
\begin{align}
\delta_1  G[x,x'] &=-
	\frac1{2\pi} G_\text N M  H^3   \frac{ \left( \eta^2 - \eta'^2 \right) \left( (\eta - \eta')^2 - 2\eta \eta' - R^2 \right) }{ \big( ( \eta - \eta' )^2 - R^2 \big)^2 } \left( 1 + \mathcal O \left[ \frac R{\eta + \eta'} , \frac R{\eta - \eta'} \right] \right) ,
	\label{PerturbedScalarGreensFunction_Leading_v2}
\end{align}
Next, we follow Poisson \cite{Poisson:2002jz} to expand eq.~\eqref{PerturbedScalarGreensFunction_Leading_v2} in powers of $\cos\gamma = \widehat x \cdot \widehat x'$, via the relation
\begin{align}
	(\cos\gamma)^{k} & = \sum_{\ell = 0}^\infty    2^{\ell + 1} \pi \left( 1 + (-1)^{k - \ell }\right)  \frac{\Gamma[k + 1] \Gamma\left[ \frac{k + \ell  }2 + 1 \right] }{\Gamma\left[\frac{k - \ell }2 + 1\right] \Gamma\left[  k + \ell + 2 \right] }   \sum_{m = -\ell}^\ell  Y^{m}_\ell[\theta, \phi] \overline{Y}^{m}_\ell[\theta',\phi'] .
\end{align}
At this point, we have recovered $\delta_1 G$ of eq. \eqref{delta1GdsFinal} within the timelike infinity limit:
\begin{align}
	\delta_1 G[x,x']
	= 2 G_\text N M  H^3  & \sum_{\ell = 0}^\infty  \sum_{m = -\ell}^\ell    (-1)^{\ell  }   \frac{ 2\ell!! }{ (2\ell + 1)!!  }  \left(\frac{ rr'}{ \eta'^2 }\right)^{\ell }    Y^{m}_\ell[\theta, \phi] Y^{m*}_\ell[\theta',\phi'] \nonumber\\
	&\qquad\qquad \times  \left( 1 + \mathcal O \left[\frac{\eta}{\eta'}, \,\, \frac r{\eta'} , \,\, \frac{r'}{\eta'}  \right] \right) .
\end{align}
\textbf{Null-Infinity} \qquad Next, let us recover the result in eq. \eqref{delta1GNull}. We start by taking the null infinity limits of $\xi$ and $\zeta$ in eqs. \eqref{I1toI6}.
\begin{align}
	\xi &= \bigg(\frac{v-u}{u+v-4\eta}\bigg) \left(1+\mathcal{O}\bigg[\frac{r'}{v}\bigg]\right) \\
	\zeta &= \bigg(\frac{v-u}{v+u}\bigg) \left(1+\mathcal{O}\bigg[\frac{r'}{v}\bigg]\right) 
\end{align}
Examining Fig. \eqref{ellipsoid} tells us, $\vec{0}$ is much closer to $\vec{x}'$ than $\vec{x}'$ is to the nearest point on the ellipsoid (namely, $L \gg r'$) if we assume that $u \gg r'$. This assumption in turn guarantees that the little error is incurred in the $1/|\vec{x}''|$ occurring within the integrals of eqs. \eqref{I1toI6} if we proceed to first expand the rest of their integrands in powers of $r'$.
{\allowdisplaybreaks\begin{align}
		I_1 &= 2 \frac{\ln[u/v]}{u-v} \left(1+\mathcal{O}[r'/u]\right)\label{I1toI6Null}\\
		I_2 &= \frac{4 \left( -u+v+\ln[u/v] \left(-2\eta+u+v\right)\right)}{(u-v)(-4\eta+u+v)} \left(1+\mathcal{O}[r'/u]\right) , \notag\\
		I_3 &= \frac{\ln\left[u(-2\eta+u)/v(-2\eta+v)\right](-4\eta+u+v)}{(u-v)(-2\eta+u+v)} \left(1+\mathcal{O}[r'/u]\right),  \notag\\
		I^{+}_{4}&= \frac{4}{u+v} \left(1+\mathcal{O}[r'/u]\right), \notag\\
		I^{-}_{4}&= \left( \frac{4 \ln[u/v]}{u-v} - \frac{4}{u+v}\right) \left(1+\mathcal{O}[r'/u]\right), \notag\\
		I^{+}_{5}&= 2\frac{(u+v-4\eta)\ln[(u-2\eta)/(v-2\eta)]}{u^{2}-v^{2}} \left(1+\mathcal{O}[r'/u]\right), \notag\\
		I^{-}_{5}&= \frac{2(-4\eta+u+v) \left(2\eta\ln[(u-2\eta)/(v-2\eta)]+\ln[u/v](u+v)\right) }{(u^{2}-v^{2})(-2\eta+u+v)}  \left(1+\mathcal{O}[r'/u]\right) , \notag\\
		I_6 &= 4\frac{\left(4\eta \tanh^{-1}[(u-v)/(-4\eta+u+v)]+u-v\right)(-4\eta+u+v)}{(u-v)(u+v)^{2}} \left(1+\mathcal{O}[r'/u]\right). \notag
\end{align}}
We subtitute all of these to obtain the perturbed Green's function $\delta_{1}G$ \eqref{PerturbedGreensFunction_dS_Position}, and proceed to take the limits $v \gg u$ and $\eta\rightarrow 0$, to obtain
\begin{align}
	\delta_1 G
	= \frac{G_{N}H^{3}M}{2\pi} \left( 1 + \frac{1}{2}\frac{v}{u} \right)  \left( 1 + \mathcal O \left[\frac{\eta}{v}, \,\, \frac u{v} , \,\, \frac{r'}{v}  \right] \right).
\end{align}
This recovers eq. \eqref{delta1GNull}.

\subsection{Behavior of Scalar Field}
In this section, we will investigate the behavior of given initial profile of scalar field \eqref{PsiInitialProfile} at the late time regime of asymptotically de Sitter spacetimes.

\textbf{Linear Propagation, $\Psi^{(0)}$} \qquad Recall the definition of $\Psi^{(0)}$ from \eqref{psipert},
\begin{align}
	\begin{split}
		\Psi^{(0)}[\eta,\eta',\vec{x}] &\equiv \int_{\Bbb R^3} \dd^{3}\vec{x}' a[\eta']^{2}  \left( \overline{G}[x,x'] \dot{\Psi}[x'] - \Psi[x']  \partial_{\eta'} \overline{G}[x,x']  \right).
	\end{split}					
\end{align}
For late-times, the delta function term in eq. \eqref{GreenFunctiondS} will not contribute to the observed scalar field $\Psi^{(0)}[\eta,\eta',\vec{x}]$. Moreover, the time derivative of $\overline{G}[x,x']$ on the second term will turn the tail of the Green function into a null signal, which will also be excluded. Additionally, $\Theta[\Delta\eta-R]=1$ at late times. Hence, we are left with
\begin{align}
	\begin{split}
		\Psi^{(0)}[\eta,\eta',\vec{x}] &\equiv a[\eta']^{2}  \int_{\Bbb R^3} \dd^{3}\vec{x}'  \frac{H^{2}}{4 \pi} \dot{\Psi}[\eta',\vec{x}'].
	\end{split}					
\end{align}
By using the initial condition of the scalar fields in eq.(\ref{PsiInitialProfile}),
\begin{align}
	\begin{split}
		\Psi^{(0)}[\eta,\eta',\vec{x}] &\equiv \frac{H^{2}}{4 \pi} a[\eta']^{2} \int_{0}^{\infty} \dd r' r'^{2} \dot{C}_{0}^{0}[r'].
		\label{Psi0}
	\end{split}	
\end{align}
In the null infinity case, \eqref{Psi0} still holds, since the step function of the unperturbed Green's function went to the same value, $\Theta[\Delta\eta-R]\approx\Theta[u]=1$. Other than that, we only need to set $\eta' \approx -v/2$,
\begin{align}
	\begin{split}
		\Psi^{(0)}_{\text{Null}}[u,v] &\equiv \frac{H^{2}}{4 \pi} a[-v/2]^{2} \int_{0}^{\infty} \dd r' r'^{2} \dot{C}_{0}^{0}[r'].
		\label{Psi0Null}
	\end{split}	
\end{align}
\textbf{Linear Propagation, $\Psi^{(0,\chi)}$} \qquad The next part is $\Psi^{(0,\chi)}$. From eq. \eqref{psipert},
\begin{align}
	\begin{split}
		\Psi^{(0,\chi)}[\eta,\eta',\vec{x}] &= \int_{\Bbb R^3} \dd^{3}\vec{x}' a^{2}[\eta'] ( -  \bar{\chi}_{00})  \left( \overline{G}[x,x']  \dot{\Psi}[x'] - \Psi[x']  \partial_{\eta'} \overline{G}[x,x']   \right).
	\end{split}					
\end{align}
The same late time considerations for $\Psi^{(0)}$ apply here, since their integral representations employ the same unperturbed retarded Green function $\overline{G}[x,x']$. We thus have
\begin{align}
	\begin{split}
		\Psi^{(0,\chi)}[\eta,\eta',\vec{x}] &= -\int_{\Bbb R^3} \dd^{3}\vec{x}' a^{2}[\eta']  \bar{\chi}_{00}[\vec{x}']  \frac{H^{2}}{4 \pi}     \sum_{\ell',m'} \dot{C}_{\ell'}^{m'}[r'] Y_{\ell'}^{m'}[\widehat{x}'],
	\end{split}					
\end{align}
where we use the initial condition of the scalar fields given in eq.(\ref{PsiInitialProfile}). 
Now, let us recall trace-reversed perturbation metric $\bar{\chi}_{00}$. Since the $r'$ integral runs over all radius, 
we will have two integration regions: outside of the mass distribution $r'>r_{\ell''}^{m''}$ and inside of mass distribution $r_{\ell''}^{m''}>r'>0$. We therefore need to invoke eq. (\ref{ChiBar00dSInside}) to deduce
\begin{align}
	\begin{split}
		\Psi^{(0,\chi)}[\eta,\eta',\vec{x}] &=4 G_\text N H^{2} \int_{0}^{\infty} \dd r' r'^{2}  a[\eta'] \sum_{\ell',m'} \left(  \frac{(-1)^{m'}}{2\ell'+1}\frac{1}{r'^{\ell'+1}}    M_{\ell'}^{m'}[r_{\ell'}^{m'}] \Theta[r'-r_{\ell'}^{m'}] \right. \\
		& \qquad\qquad\qquad \left. +   \frac{(-1)^{m'}}{2\ell'+1}   \left\{ \frac{ M_{\ell'}^{m'}[r']}{r'^{\ell'+1}} + N_{\ell'}^{m'}[r'] r'^{\ell'}\right\}  \Theta[r_{\ell'}^{m'}-r'] \right)   \dot{C}_{\ell'}^{m'}[r'] ,
		\label{Psi0chi}
	\end{split}					
\end{align}
where the first and second terms respectively describe the contribution from the initial scalar field outside (eq. \eqref{ChiBar00dSOutside}) and inside (eq. \eqref{ChiBar00dSInside}) the central mass, with $M_{\ell'}^{m'}[r']$ and $N_{\ell'}^{m'}[r']$ already defined in eq. \eqref{InternalMultipoles}. Like its cousin in eq. \eqref{Psi0}, the result in eq. \eqref{Psi0chi} is a constant with respect to $(\eta,\vec{x})$; and therefore its null infinity expression remains the same, except $\eta'$ may be identified with $-v/2$.

While only the initial scalar field velocity's monopole contributed to the central mass independent linear propagation result in eq. \eqref{Psi0}; we see from eq. \eqref{Psi0chi}, all the initial velocity's multipoles $\{ \ell',m' \}$ do contribute once the gravitational potential of the central mass is included, because these $\{ \ell',m' \}$ are now tied to the mass multipoles $\{ \ell'',m'' \}$ in a one-to-one manner, namely $\ell' = \ell''$ and $m' = m''$. Moreover, both equations \eqref{Psi0} and \eqref{Psi0chi} describe a final scalar field that does not decay back to zero amplitude, but instead asymptotes to a constant. In addition, we may compare equations \eqref{Psi0} and \eqref{Psi0chi} to the results obtained by BCLK \cite{Brady:1996za} and BCLP \cite{Brady:1999wd}, where they dealt with a spherically symmetric black hole in de Sitter spacetime. Like them, we find the final scalar field to scale as $H^2$; but unlike them, because of our arbitrary matter distribution, we uncovered sensitivity to both external and internal multipoles of all orders. Even though BCLK and BCLP did not name it as such, they found -- as we did here -- their massless scalar field to develop a memory at late times.

\textbf{Non-Linear Propagation, $\Psi^{(1)}$} \qquad Now, let us find out how the non-linear parts contribute to the scalar signal. Here, the propagation is governed by the first order perturbed retarded Green's function. Recall the definition of $\Psi^{(1)}$ from \eqref{psipert},
\begin{align}
	\Psi^{(1)}[\eta,\eta',\vec{x}]	= \int_{\Bbb R^3} \dd^{3}\vec{x}' a[\eta']^{2} \left( \delta_{1}G[x,x'] \dot{\Psi}[x'] - \Psi[x']  \partial_{\eta'} \delta_{1}G[x,x'] \right)	.
\end{align}
The perturbed Green's function for timelike infinity observer has been computed in eq. \eqref{delta1GdsFinal}. With the initial profile in eq. \eqref{PsiInitialProfile},
\begin{align}
	\label{Psi1final}
	&\Psi^{(1)}[\eta,\eta',\vec{x}]	
	= a^{2}[\eta'] M \GN H^{3} \int_{0}^{\infty} \dd r' r'^{2} \Bigg\{ \frac{\dot{C}_{0}^{0}[r']}{4 \pi^{3/2}}  \\
	&\qquad\qquad
	+    \sum_{\ell'=1,m'}  \left( \dot{C}_{\ell'}^{m'}[r']  - \frac{2\ell}{\eta'} C_{\ell'}^{m'}[r']  \right) Y_{\ell'}^{m'}[\widehat{x}] \left(\frac{ rr'}{ \eta'^2 }\right)^{\ell'}
	\frac{2(2\ell')!!(-1)^{\ell'}}{(2\ell' + 1)!!} \Bigg\}  \left( 1 + \mathcal{O}\left[\frac{\eta}{\eta'},\left(\frac{r}{\eta'}\right)^{2},\left(\frac{r'}{\eta'}\right)^{2}\right]  \right) . \notag
\end{align}
Observe that, because only the mass monopole term contributes to $\delta_1 G$ in eq. \eqref{delta1GdsFinal}, there is an angular momentum conservation at play here: the initial $\ell$th multipole of the scalar field at $\eta'$ propagates forward in time into a pure $\ell$-multipole at late times $\eta \to 0$.

In the late retarded time null infinity limit, the perturbed Green's function is given in eq.\eqref{delta1GNull}. By using the same initial profile as timelike infinity observer scenario \eqref{PsiInitialProfile},
\begin{align}
	\Psi^{(1)}_{\text{Null}}[u,v] = \frac{G_{N}H^{3}M }{(4\pi)^{3/2}}\int_{0}^{\infty} \dd r' r'^{2} a[-v/2]^{2} \left( \left(\frac{v}{u}+2\right)  \dot{C}^{0}_{0}[r'] + \frac{C^{0}_{0}[r']}{ u} \left(1-\frac{v}{u}\right)  \right)	.
\end{align}
For this calculation, we have only computed the monopole solution.

{\bf Comparison to Static Patch Results} \qquad The line element of de Sitter in static patch coordinates is given by
\begin{align}
	\dd s^{2}= (1-H^{2}r_{s}^{2}) \dd t_{s}^{2}-(1-H^{2}r_{s}^{2})^{-1} \dd r_{s}^{2}-r^{2}_{s} \dd\Omega^{2}.
\end{align}
The transformation rules from flat slicing coordinates $(\eta,r)$ in eq. \eqref{PerturbedGeometry} to static patch coordinates $(t_{s},r_{s})$ is
\begin{align}
	r = \frac{r_{s} \exp[-H t_{s}] }{\sqrt{1-H^2 r_{s}^2}}, \qquad \text{and} \qquad
	\eta = -\frac{\exp[-H t_{s}]}{H \sqrt{1-H^2 r_{s}^2}}.
\end{align}
By employing these transformation rules to $r^{\ell}$ in our solution (\ref{Psi1final}), the $\ell$th multipole of the scalar field becomes
\begin{align}
	\Psi^{(1)} \sim e^{-\ell H t_{s}}.
	\label{PsiBCLP}
\end{align}
This decay law was first obtained by BCLP \cite{Brady:1999wd}.

We also highlight that the late time assumption that allowed us to simplify the Green's function expression to that in eq. \eqref{delta1GdS(46)}, while made in the flat slicing coordinates of eq. \eqref{PerturbedGeometry}, continues to hold even in the static patch of de Sitter spacetime. This is because $\eta/\eta' \approx \exp[-H(t_s-t_s')] \to 0$ as $t_s \to \infty$.

\section{Summary and Future Work}
In this paper, we have calculated the behavior of a scalar field in the late-time regime of perturbed de Sitter spacetime, where the perturbation is generated by a localized mass distribution. We performed the calculation in a (fictitious) frequency space that amounted to replacing delta functions with their integral representations. This allowed us to implement a spherical harmonic decomposition early on in the computation. We tested the method in \S \eqref{Section_MinkowskiFrequencySpace} by recovering a portion of Poisson's asymptotically flat results in \cite{Poisson:2002jz}.

By taking the leading order terms of the $\eta/\eta' \to 0$ and $-r'/\eta' < -r/\eta' \ll 1$ expansion at the early stage of calculation, we recognized three contributions to the observed scalar field at $(\eta,\vec{x})$ in the timelike infinity limit. The first term is the linear part of the scalar field propagation itself, coming from the tail of the de Sitter massless scalar Green's function. The second is just the linear part with a first order metric perturbation, where the initial scalar field is sensitive to both the internal and external mass multipoles. The first and second terms yield linear scalar memory, as $\eta/\eta' \to 0$ at fixed $r$: $\Psi$ does not decay back to zero but to a spacetime constant proportional to $H^2$. The third term is the non-linear part of the scalar field propagation, governed by the perturbed Green's function. Here, the scalar interacts with the gravitational potential of the central mass through scattering off the intersection of the future lightcone of the initial scalar profile and past lightcone of the observer. There is a nonlinear scalar memory effect like the linear case, where the scalar decays to a spacetime constant that scales as $(\GN M) H$ relative to the linear memory, where $M$ is the total mass of the matter distribution. Moreover, this nonlinear scalar memory does not appear to be sensitive at all to the higher mass multipoles. We also provided an analytic understanding of the decay law first uncovered by BCLK \cite{Brady:1996za} for the higher multipoles of the final scalar field: in static patch coordinate time $t_s$, the $\ell$th moment decays in time as $\Psi^{(1)} \sim H^{3} \exp(-\ell H t_s)$.

We also examined the monopole portion of the same scalar field signal but at null infinity $v \to -2\eta' \gg u$ and late retarded times $u \gg r'$. It exhibited a novel feature compared to its timelike infinity cousin. Specifically, it contains an enhancement that scales as $v/(2u)$ relative to the nonlinear spacetime-constant memory. In the future, we hope to further investigate the physical origin(s) of this term, as well as its potential implications for the convergence of classical perturbation theory on de Sitter backgrounds.

To further develop the work here, we may consider more general matter distributions or perhaps dynamical astrophysical systems. It could be of physical interest to extend our late time analysis to that of intermediate times, to understand the full causal structure of the scalar signal. We also wish to extend the analysis here to that of gravitational waves -- and ask: what sort of nonlinear gravitational memories are there in asymptotically de Sitter spacetimes?

\section{Acknowledgements}

We were supported by the Ministry of Science and Technology of the R.O.C. under the grant 106-2112-M-008-024-MY3. YWL is currently supported by the Ministry of Science and Technology of the R.O.C. under Project No. MOST 109-2811-M-007-514

\appendix

\section{The Multipoles}
\label{Section_Multipoles}

\subsection{Evaluation of Double Sum}
\label{Section_DoubleSum}

In this paper, we encountered the following double summation that appeared in the spherical harmonic expansion of the perturbed scalar Green's function $\delta_1 G$ in both Minkowski and de Sitter spacetimes:
\begin{align}
	S[\ell,\ell',\ell''] \equiv \sum_{n = 0}^\ell  \sum_{n' = 0}^{\ell'}  \frac{ (-1)^{n + n'} }{(\ell'' + n + n')!}  \frac{(\ell+n)!}{ n! (\ell-n)!} \frac{(\ell' + n')!}{ n'! (\ell' - n')!} ,
	\label{DoubleSum}
\end{align}
We may first perform the $n'$ summation in eq.~\eqref{DoubleSum} to obtain
{\allowdisplaybreaks
	\begin{align}
		S[\ell,\ell',\ell'']	& = \sum_{n = 0}^\ell  (-1)^n  \frac{ \Gamma[\ell+n + 1] \Gamma[ \ell'' + n] }{ \Gamma[n+1] \Gamma[ \ell - n + 1]   \Gamma[ \ell'' + \ell' + n + 1 ] \Gamma[ \ell'' - \ell' + n  ] }   , \qquad \text{Re}[\ell''] > 0 ,
		\label{FirstSum}
\end{align}}%
where we have used the following limit to replace the $(-1)^{n'}$ factor in eq.~\eqref{DoubleSum},
\begin{align}
	\frac{\Gamma[ - \ell' + n']}{\Gamma[-\ell']}  =  \frac{ (-1)^{n'} \Gamma[\ell'+1] }{ \Gamma[\ell' - n'+ 1]} , \qquad 0 \leq n' \leq \ell' .
	\label{Ratio_GammaFunctions}
\end{align}
as well as the identity
\begin{align}
	_2F_1[ - \ell' , \ell' + 1 ; \ell'' + n + 1 ; 1 ]  = \frac{ \Gamma[\ell'' + n + 1] \Gamma[ \ell'' + n] }{ \Gamma[ \ell'' + \ell' + n + 1 ] \Gamma[ \ell'' - \ell' + n  ]  } , \qquad \text{Re}[\ell'' + n] > 0 .
\end{align}
Strictly speaking, eq.~\eqref{FirstSum} does not hold for the monopole $\ell'' = 0$ case. Nevertheless, we may proceed to compute the $n$-summation assuming $\ell''>0$, followed by taking the $\ell'' \to 0$ limit.

To perform the $n$-summation, we may exploit the replacement \eqref{Ratio_GammaFunctions} again (with $n'\to n$ and $\ell' \to \ell$) to reach
{\allowdisplaybreaks
	\begin{align}
		S[\ell,\ell', \ell''] 		& =   \frac{  2^{1 - 2 \ell'' } \pi \Gamma[\ell''] }{ \Gamma\left[ \frac12 (\ell'' - \ell - \ell' ) \right] \Gamma\left[ \frac12 ( \ell'' + \ell - \ell' + 1) \right] \Gamma\left[ \frac12 ( \ell'' - \ell + \ell' + 1 )  \right] \Gamma\left[ \frac12 ( \ell'' + \ell + \ell' + 2 ) \right]  } ,
		\label{SecondSum}
\end{align}}%
for $\text{Re}[\ell''] > 0$, where we have employed one of Whipple's identities (see [DLMF, Eq. 16.4.7]\cite{NIST}):
\begin{align}
	&_3F_2  [a, 1 - a, c \, ; d , 2c - d + 1; 1]  \\
	&\qquad = \, \frac{  2^{1 - 2 c} \pi \Gamma[d] \Gamma[2c - d + 1] }{ \Gamma\left[ c + \frac12 (a - d + 1) \right] \Gamma\left[ c + 1 - \frac12(a + d) \right] \Gamma\left[ \frac12 (a + d)  \right] \Gamma\left[ \frac12 ( d - a + 1) \right]  } , \qquad \text{Re}[c] > 0 . \notag
\end{align}

\subsection{Constraints on Multipole Indices from Parity}
\label{Section_ParityArgument}
Up to this point, the only constraint on the multipole indices occuring within the $\delta_{1}G$ comes from the usual rules of angular momentum addition. In this section, we will discover another set of constraints from parity considerations.

In $\delta_{1}G$, we have three multipoles $\ell$, $\ell'$, and $\ell''$ that correspond respectively to the multipoles of the observed scalar field, initial profile of the scalar field, and the mass distribution. Let us recall the structure of $\delta_{1}G$, up the scale factor for de Sitter case,
\begin{align}
	\begin{split}
		\delta_{1}G[t,t';\vec{x},\vec{x}'] &\sim \int \dd^{4}x'' \partial_{t''}\left(\frac{\delta[t-t''-|\vec{x}-\vec{x}''|]}{4\pi |\vec{x}-\vec{x}''|} \right) \int  \dd^{3}\vec{x}''' \frac{\rho[\vec{x}''']}{|\vec{x}'' - \vec{x}'''|} \partial_{t''}\left(\frac{\delta[t''-t'-|\vec{x}'' - \vec{x}'|]}{4\pi |\vec{x}'' - \vec{x}'|} \right)
	\end{split}					
\end{align}
Upon the parity flip $\vec{x} \rightarrow -\vec{x}$ and $\vec{x}' \rightarrow -\vec{x}'$, followed a similar change in the integration variables,
\begin{align}
&\delta_{1}G[t,t';-\vec{x},-\vec{x}'] \nonumber\\
&\qquad
\sim \int \dd^{4}x'' \partial_{t''}\left(\frac{\delta[t-t''-|(-\vec{x}) -\vec{x}''|]}{4\pi |(-\vec{x})-\vec{x}''|} \right) \int  \dd^{3}\vec{x}''' \frac{\rho[\vec{x}''']}{|\vec{x}'' - \vec{x}'''|} \partial_{t''}\left(\frac{\delta[t''-t'-|\vec{x}'' - (-\vec{x}')|]}{4\pi |\vec{x}'' - (-\vec{x}')|} \right)
		\label{306} \\
&\qquad
= \int  \dd^{4}x'' \partial_{t''}\left(\frac{\delta[t-t''-|\vec{x} -\vec{x}''|]}{4\pi |\vec{x}-\vec{x}''|} \right) \int  \dd^{3}\vec{x}''' \frac{\rho[-\vec{x}''']}{|\vec{x}'' - \vec{x}'''|} \partial_{t''}\left(\frac{\delta[t''-t'-|\vec{x}'' - \vec{x}'|]}{4\pi |\vec{x}'' - \vec{x}'|} \right)
		\label{307}
\end{align}
The spherical harmonics dependence of each multipole in eq. (\ref{306}) is
\begin{align*}
	\begin{split}
		Y_{\ell}^{m}[\widehat{x}] 	\bar{Y}_{\ell}^{m}[\widehat{x}'']  Y_{\ell'}^{m'}[\widehat{x}''] \bar{Y}_{\ell'}^{m'}[\widehat{x}'] 	Y_{\ell''}^{m''}[\widehat{x}'''] (-)^{\ell+\ell'},
	\end{split}					
\end{align*}
whereas the spherical harmonic dependence in eq. (\ref{307}) is
\begin{align*}
	\begin{split}
		Y_{\ell}^{m}[\widehat{x}] 	\bar{Y}_{\ell}^{m}[\widehat{x}'']  Y_{\ell'}^{m'}[\widehat{x}''] \bar{Y}_{\ell'}^{m'}[\widehat{x}'] 	Y_{\ell''}^{m''}[\widehat{x}'''] (-)^{\ell''}.
	\end{split}					
\end{align*}
Since spherical harmonics are linearly independent, equations (\ref{306}) and (\ref{307}) should in fact be the same:
\begin{align}
	\begin{split}
		\ell+\ell'=\ell''+2 q, \qquad q \in \mathbb{Z} .
	\end{split}					
\end{align}

\section{Contributions to $\delta_{1}G$ in de Sitter Background}
\label{Section_ABC}

\subsection{Spherical Harmonic Decomposition of $\exp[i\omega R]/R^{2}$}
\label{SphericalHarmonicDecomp_ExpDivR2}

As we shall witness below, the spherical harmonic decomposition of $\exp[i\omega R]/R^{2}$, where $R \equiv |\vec{x}-\vec{x}''|$, is needed in the computation of $\delta_1 G$ in de Sitter. First re-express it as
\begin{align}
	\begin{split}
		\frac{e^{i \omega R}}{4\pi R^{2}} &= i \int^{\omega}_{0} \dd\bar{\omega}  \frac{e^{i \bar{\omega} R}}{4\pi R} +  \frac{1}{4\pi R^{2}}.
		\label{eiomegaR/R^2}
	\end{split}					
\end{align}
The first term can be computed by integrating (\ref{GFreqFlat1}),
\begin{align}
	\begin{split}
		i \int^{\omega}_{0} \dd\bar{\omega}  \frac{e^{i \bar{\omega} R}}{4\pi R} &= \left(   \sum_{\ell,m}  \frac{(-i r)^{\ell}}{r''(2\ell+1)!!} \sum_{s=0}^{\ell} \frac{i^{s}}{s! (2  r'')^{s}} \frac{(\ell+s)!}{(\ell-s)!}  Y_{\ell}^{m}[\widehat{x}]  \bar{Y}_{\ell}^{m}[\widehat{x}'']   \sum_{k=0}^{\ell-s} \frac{C[\ell-s,k] e^{i\omega r''}k!}{( \omega)^{-\ell+s+k} (-i)^{k} }     r''^{-1-k} \right.\\
		&\left. \hphantom{aaaa} - \sum_{\ell,m} \left(\frac{r}{r''} \right)^{\ell}  \frac{ 2^{\ell} \ell!  Y_{\ell}^{m}[\widehat{x}]  \bar{Y}_{\ell}^{m}[\widehat{x}'']}{r''^{2}(2\ell+1)!!} \right)  \left(1+\mathcal{O}[(r/r'')^{2}]\right) .
		\label{eiomegaR/R^2(1)}	
	\end{split}					
\end{align}
For the second term,  we refer to Dixon and Lacroix work \cite{Dixon_1973}. They tell us
\begin{align}
	\begin{split}
		\frac{1}{4\pi|\vec{x}-\vec{x}''|^{2}} &= \frac{1}{4\pi} \sum_{\ell=0}^{\infty} \frac{2\ell+1}{2rr''} Q_{\ell}\left[\frac{1+(r/r'')^{2}}{2(r/r'')} \right] P_{\ell}[\widehat{x}\cdot \widehat{x}'']  . \\
	\end{split}					
\end{align}
We can approximate $(r/r'')\rightarrow 0$. In doing so, we need to exploit the behavior of $ Q_{\ell}\left[\frac{1+(r/r'')^{2}}{2(r/r'')} \right]$ at infinity, which gives us
\begin{align}
	\begin{split}
		\frac{1}{4\pi|\vec{x}-\vec{x}''|^{2}} &\approx   \sum_{\ell,m} \frac{\ell!}{ r''^{2}} \left(\frac{r}{r''} \right)^{\ell} \frac{2^{\ell}}{(2\ell+1)!!}  Y_{\ell}^{m}[\widehat{x}]  \bar{Y}_{\ell}^{m}[\widehat{x}''] .
		\label{eiomegaR/R^2(2)}
	\end{split}					
\end{align}
By combining equations (\ref{eiomegaR/R^2(1)}) and (\ref{eiomegaR/R^2(2)}),
\begin{align}
	\label{expDivR2}
	\begin{split}
		\frac{e^{i \omega R}}{4\pi R^{2}} &\approx   \sum_{\ell,m}  \frac{(-i r)^{\ell}}{r''(2\ell+1)!!} \sum_{s=0}^{\ell} \frac{i^{s}}{s! (2  r'')^{s}} \frac{(\ell+s)!}{(\ell-s)!}  Y_{\ell}^{m}[\widehat{x}]  \bar{Y}_{\ell}^{m}[\widehat{x}'']   \sum_{k=0}^{\ell-s} \frac{C[\ell-s,k] e^{i\omega r''}k!}{( \omega)^{-\ell+s+k} (-i)^{k} }     r''^{-1-k}  \left(1+\mathcal{O}[(r/r'')^{2}] \right) .
	\end{split}					
\end{align}
Observe that (\ref{eiomegaR/R^2(2)}) canceled the second term of (\ref{eiomegaR/R^2(1)}).

\subsection{$A_{a}$}
Recall the definition of $A_{a}$ from \eqref{Aa}. We can collapse one delta function by simply integrating $\eta''$,
\begin{align}
	\begin{split}
		A_{a} &\equiv \int_{-\infty}^{0}  \dd\eta'' \eta''^{a}  \frac{\delta[\eta-\eta''-|\vec{x}-\vec{x}''|]}{4 \pi |\vec{x}-\vec{x}''|} \frac{\delta[\eta''-\eta'-|\vec{x}''-\vec{x}'|]}{4 \pi |\vec{x}''-\vec{x}'|}\\
		&= \left( \eta-|\vec{x}-\vec{x}''| \right)^{a}  \frac{\delta[\eta-\eta'-|\vec{x}''-\vec{x}'|-|\vec{x}-\vec{x}''|]}{4 \pi |\vec{x}-\vec{x}''|4 \pi |\vec{x}''-\vec{x}'|}  ,
	\end{split}					
\end{align}
where $a=-1,0,1$. Different value of $a$ may lead to different results.
\subsubsection{$A_{0}$}
For $a=0$, we have
\begin{align}
	\begin{split}
		A_{0} \equiv \int_{-\infty}^{\infty} \frac{ \dd\omega}{2\pi}  e^{-i \omega(\eta-\eta')} \frac{e^{i \omega(|\vec{x}-\vec{x}''|)}}{4 \pi |\vec{x}-\vec{x}''|} \frac{e^{i \omega(|\vec{x}''-\vec{x}'|)}}{4 \pi |\vec{x}''-\vec{x}'|}.
	\end{split}					
\end{align}

We can impose multipole expansions of $\frac{e^{i \omega(|\vec{x}-\vec{x}''|)}}{4 \pi |\vec{x}-\vec{x}''|}$ and $\frac{e^{i \omega(|\vec{x}''-\vec{x}'|)}}{4 \pi |\vec{x}''-\vec{x}'|}$ with $\omega$ as the frequency and spherical harmonics as basis functions. Here we take the small argument limit of spherical bessel $j_{\ell}[\omega r]$ and $j_{\ell}[\omega r']$ since they are small quantities in the time-like infinity case. After we evaluate the $\omega$ integral, we have
\begin{align}
		A_{0}  &=  i  \sum_{\ell,m} \frac{(r)^{\ell}}{(2\ell+1)!!}  \frac{(-i)^{\ell+1}}{ r''} \sum_{s=0}^{\ell} \frac{i^{s}}{s! (2  r'')^{s}} \frac{(\ell+s)!}{(\ell-s)!}Y_{\ell}^{m}[\widehat{x}]  \bar{Y}_{\ell}^{m}[\widehat{x}'']\\
		&\hphantom{aaa} \times i  \sum_{\ell',m'} \frac{( r')^{\ell'}}{(2\ell' + 1)!!}   \frac{(-i)^{\ell'+1}}{ r''} \sum_{s'=0}^{\ell'} \frac{i^{s'}}{s'! (2  r'')^{s'}} \frac{(\ell'+s')!}{(\ell'-s')!}Y_{\ell'}^{m'}[\widehat{x}'']  \bar{Y}_{\ell'}^{m'}[\widehat{x}']
		(i \partial_{\eta})^{\ell-s+\ell'-s'} \delta[(\eta-\eta')-2r''] . \notag
		\label{A0} 				
\end{align}
By including the $r''$ integral and $r''^{-\ell+1}$, and referring to the $\delta_{1}G$ expression in eq.(\ref{delta1GdS(46)}), we can collapse the delta function that contains $r''$ into $\eta-\eta'$ and evaluate the $\eta'$ derivative
\begin{align}
	\int  \dd r'' r''^{-\ell''+1} A_{0} &= -   \sum_{\ell,m} \frac{(r)^{\ell}}{(2\ell+1)!!}    \sum_{s=0}^{\ell} \frac{1}{s! } \frac{(\ell+s)!}{(\ell-s)!}Y_{\ell}^{m}[\widehat{x}]  \bar{Y}_{\ell}^{m}[\widehat{x}'']  \sum_{\ell',m'} \frac{( r')^{\ell'}}{(2\ell' + 1)!!}     \sum_{s'=0}^{\ell'} \frac{1}{s'! } \frac{(\ell'+s')!}{(\ell'-s')!}Y_{\ell'}^{m'}[\widehat{x}'']  \bar{Y}_{\ell'}^{m'}[\widehat{x}'] \notag \\
	&\hphantom{aaaaaa} \times  2''^{\ell''} (-1)^{-\ell -\ell'+s+s'} (\eta - \eta')^{-\ell -\ell'-\ell''-1} \frac{\Gamma [\ell +\ell'+\ell''+1]
	}{\Gamma [s+s'+\ell''+1]}		 .
\end{align}

\subsubsection{$A_{1}$}
For $a=1$, we have
\begin{align}
	A_{1} &\equiv  \int_{-\infty}^{\infty} \frac{ \dd\omega}{2\pi}  e^{-i \omega(\eta-\eta')} (\eta-|\vec{x}-\vec{x}''|)  \frac{e^{i \omega(|\vec{x}-\vec{x}''|)}}{4 \pi |\vec{x}-\vec{x}''|} \frac{e^{i \omega(|\vec{x}''-\vec{x}'|)}}{4 \pi |\vec{x}''-\vec{x}'|} , \notag \\
	&= \int_{-\infty}^{\infty} \frac{ \dd\omega}{2\pi}  e^{-i \omega(\eta-\eta')} \eta  \frac{e^{i \omega(|\vec{x}-\vec{x}''|)}}{4 \pi |\vec{x}-\vec{x}''|} \frac{e^{i \omega(|\vec{x}''-\vec{x}'|)}}{4 \pi |\vec{x}''-\vec{x}'|}  - \int_{-\infty}^{\infty} \frac{ \dd\omega}{2\pi}  e^{-i \omega(\eta-\eta')}   \frac{e^{i \omega(|\vec{x}-\vec{x}''|)}}{4 \pi } \frac{e^{i \omega(|\vec{x}''-\vec{x}'|)}}{4 \pi |\vec{x}''-\vec{x}'|} , \notag \\
	& \equiv \eta A_{0} + A_{1,2} .
\end{align}
The first term is exactly $A_{0}$ multiplied by $\eta$; while the second term $A_{1,2}$ is unknown at this point. Therefore, we need to compute the latter separately. Note that the spherical harmonic expansion of $\exp[i \omega(|\vec{x}-\vec{x}''|)]/4 \pi $ can be obtained from that of $\exp[i \omega(|\vec{x}''-\vec{x}'|)]/4 \pi |\vec{x}''-\vec{x}'|$ (which is a standard result) via differentiation, namely
\begin{align}
	\label{DerivativeOfExpROverR}
	-i \partial_\omega \frac{\exp[i \omega(|\vec{x}''-\vec{x}'|)]}{4 \pi |\vec{x}''-\vec{x}'|}
	= \frac{\exp[i \omega(|\vec{x}-\vec{x}''|)]}{4 \pi} .
\end{align}
This leads us to
\begin{align}
	A_{1,2}&  = - \int_{-\infty}^{\infty} \frac{ \dd\omega}{2\pi}  e^{-i \omega(\eta-\eta')}  \sum_{\ell,m}\frac{ r'' r \omega ^2 }{2 \ell +1} \left(h_{\ell	-1}^{(1)}[r'' \omega ] j_{\ell -1}[r \omega ]-h_{\ell +1}^{(1)}[r'' \omega ] j_{\ell	+1}[r \omega ] \right)  Y_{\ell}^{m}[\widehat{x}]  \bar{Y}_{\ell}^{m}[\widehat{x}''] \notag\\
	&\hphantom{aaaaaaaaaaaaa} \times i \omega \sum_{\ell',m'}j_{\ell'}[\omega r']h^{(1)}_{\ell'}[\omega r''] Y_{\ell'}^{m'}[\widehat{x}'']  \bar{Y}_{\ell'}^{m'}[\widehat{x}'] .
	\label{A12form}
\end{align}
We cannot take small argument limit of $j_{-1}[\omega r]$ directly. What we should do instead is to do the $\ell=0$ separately. By using a certain recursion relation of the spherical Bessel function, we have
\begin{align}
	A_{1,2}^{\ell=0}&= - \int_{-\infty}^{\infty} \frac{ \dd\omega}{2\pi}  e^{-i \omega(\eta-\eta')}    r'' r \omega ^2  \left(\left(-h_{1}^{(1)}[r'' \omega ] + \frac{1}{\omega r''}h_{0}^{(1)}[r'' \omega ] \right) \left(-j_{1}[r \omega ] + \frac{1}{\omega r} j_{0}[r \omega ]  \right)-h_{ 1}^{(1)}[r'' \omega ] j_{1}[r \omega ] \right) \notag \\
	&\hphantom{aaaaaaaaaaaaa} \times Y_{0}^{0}[\widehat{x}]  \bar{Y}_{0}^{0}[\widehat{x}''] i \omega \sum_{\ell',m'}j_{\ell'}[\omega r']h^{(1)}_{\ell'}[\omega r''] Y_{\ell'}^{m'}[\widehat{x}'']  \bar{Y}_{\ell'}^{m'}[\widehat{x}']\ \delta_{\ell,0}		.
\end{align}
The same procedure as the previous calculation can be used at this level. Take the small argument limit of $j_{\ell}[\omega r]$, evaluate the $\omega$ integral, and the include $r''$ integral to obtain
\begin{align}
	\int_{0}^{\infty}  \dd r'' r''^{-\ell''+1} A_{1,2}^{\ell=0} &= -      \frac{ \delta_{\ell,0} }{4\pi} \sum_{\ell',m'}  \frac{( r')^{\ell'}(\eta -\eta')^{-\ell'-\ell''-2}}{(2\ell'+1)!!}   \sum_{s'=0}^{\ell'} \frac{1}{s'! } \frac{(\ell'+s')!}{(\ell'-s')!} Y_{\ell'}^{m'}[\widehat{x}'']  \bar{Y}_{\ell'}^{m'}[\widehat{x}'] \notag  \\
	&\hphantom{aaaaa} \times 2^{\ell''-1}   (-1)^{-\ell'+s'} \frac{ (\ell'+\ell'')! }{ (s'+\ell'')! }   \left( \frac{2 r^{2}}{3}     (\ell'+\ell''+1)   +   (\eta -\eta')^{2}  \right) 		.
\end{align}
The first term is supressed by $(r/\eta')^{2}$, while the second term is the leading term which contributes to the overall power of $\eta'$.

For $A_{1,2}$ with $\ell \geq 1$ in eq.(\ref{A12form}),
\begin{align}
		A_{1,2}^{\ell \geq 1}&\equiv  - \int_{-\infty}^{\infty} \frac{ \dd\omega}{2\pi}  e^{-i \omega(\eta-\eta')}  \sum_{\ell=1,m}\frac{ r'' r \omega ^2 }{2 \ell +1} \left(h_{\ell	-1}^{(1)}[r'' \omega ] j_{\ell -1}[r \omega ]-h_{\ell +1}^{(1)}[r'' \omega ] j_{\ell	+1}[r \omega ] \right)  Y_{\ell}^{m}[\widehat{x}]  \bar{Y}_{\ell}^{m}[\widehat{x}''] \notag \\
		&\hphantom{aaaaaaaaaaaaa} \times i \omega \sum_{\ell',m'}j_{\ell'}[\omega r']h^{(1)}_{\ell'}[\omega r''] Y_{\ell'}^{m'}[\widehat{x}'']  \bar{Y}_{\ell'}^{m'}[\widehat{x}'] .	
\end{align}
This can be computed with the same procedure as previous calculation. Here we can directly take small argument limit of $j_{\ell}[\omega r]$. By including the $r''$ integral, its solution is
\begin{align}
	\int_{0}^{\infty}  \dd r'' r''^{-\ell''+1}  A_{1,2}^{\ell \geq 1}&= -\sum_{\ell=1,m} \sum_{\ell',m'} \sum_{s'=0}^{\ell'}  \frac{ 2^{\ell''-1} r^{\ell } r'^{\ell'}
		(-1)^{-s'+\ell + \ell'} (\eta - \eta')^{-\ell''-\ell -\ell'-2} }{(2 \ell +1) (2 \ell'+1)!!  s'! } \frac{ (\ell'+s')!}{(\ell'-s')!} \\
	& \times \left(\sum_{s=0}^{\ell-1}  \frac{(-)^{s} (\eta-\eta')^{2} \Gamma (s+\ell )  \Gamma (\ell''+\ell +\ell')}{(2 \ell -1)\text{!!} \Gamma (\ell -s) \Gamma[\ell''+s+s'] s!}- \sum_{s=0}^{\ell+1}  \frac{(-)^{s} r^2 \Gamma (s+\ell +2) 	\Gamma (\ell''+\ell +\ell'+2)}{(2 \ell +3)\text{!!}  \Gamma
		[2-s+\ell ] \Gamma[\ell''+s+s'] s!}\right).	\notag
\end{align}

\subsubsection{$A_{-1}$}
For $a=-1$, we have
\begin{align}
	\begin{split}
		A_{-1} &\equiv \int_{-\infty}^{\infty} \frac{ \dd \omega}{2\pi} \frac{ e^{-i \omega(\eta-\eta')}}{ \eta-|\vec{x}-\vec{x}''| } \frac{e^{i \omega(|\vec{x}-\vec{x}''|)}}{4\pi|\vec{x}-\vec{x}''|} \frac{e^{i \omega(|\vec{x}''-\vec{x}'|)}}{4\pi|\vec{x}''-\vec{x}'|} .
	\end{split}					
\end{align}
At late time $\eta \rightarrow 0$, as long as $|\vec{x}-\vec{x}''| \neq 0$, we can expand $(1-\frac{\eta }{|\vec{x}-\vec{x}''|})^{-1}$ around $\eta = 0$. Up to the zeroth order, we have
\begin{align}
	\begin{split}
		A_{-1} 		&\approx \int_{-\infty}^{\infty} \frac{ \dd\omega}{2\pi} \frac{ e^{-i \omega(\eta-\eta')}}{-|\vec{x}-\vec{x}''|  } \frac{e^{i \omega(|\vec{x}-\vec{x}''|)}}{4\pi|\vec{x}-\vec{x}''|} \frac{e^{i \omega(|\vec{x}''-\vec{x}'|)}}{4\pi|\vec{x}''-\vec{x}'|}  \left(1 + \mathcal{O}[\frac{\eta }{|\vec{x}-\vec{x}''|} ] \right) .
		\label{n-1}
	\end{split}					
\end{align}
Let's employ spherical harmonic decomposition on this expression. Notice that the multipole expansion of $\exp[i\omega R]/R^{2}$ is non-trivial -- see appendix \eqref{SphericalHarmonicDecomp_ExpDivR2}.
\begin{align}
	A_{-1} 	&\approx -\int_{-\infty}^{\infty} \frac{ \dd\omega}{2\pi}  e^{-i \omega(\eta-\eta')}  \   \sum_{\ell,m} \frac{(-r\omega)^{\ell }}{(2 \ell+1)!!} \frac{e^{i  \omega r'' }}{r''^{2}} \sum_{s=0}^{\ell} \frac{i^{s+\ell }}{2^{s}s!} \frac{(\ell+s)!}{(\ell-s)!} \sum_{k=0}^{\ell-s}   (-i)^{-k} 	k! (\omega r'')^{-k-s }  C[\ell-s,k] Y_{\ell}^{m}[\widehat{x}]  \bar{Y}_{\ell}^{m}[\widehat{x}'']  \notag \\
	&\hphantom{aaaaaaaaaaaa} \times i \omega \ \sum_{\ell',m'}   \frac{(\omega r')^{\ell'}}{(2\ell'+1)!!}   \frac{e^{i \omega r''}}{\omega r''} \sum_{s'=0}^{\ell'} \frac{i^{s'}(-i)^{\ell'+1}}{s'! (2 \omega r'')^{s'}} \frac{(\ell'+s')!}{(\ell'-s')!}  Y_{\ell'}^{m'}[\widehat{x}'']  \bar{Y}_{\ell'}^{m'}[\widehat{x}'] .
	\label{omega}		
\end{align}
By following the same scenario as previous calculation, and including the $r''$ integral, we have
\begin{align}
	\int_{0}^{\infty}  \dd r'' r''^{-\ell''+1} A_{-1} 	&= -      \sum_{\ell,m} \frac{(r)^{\ell }}{(2 \ell+1)!!} \sum_{s=0}^{\ell} \frac{1}{s!} \frac{(\ell+s)!}{(\ell-s)!} \sum_{k=0}^{\ell-s}  	k!   C[\ell-s,k] Y_{\ell}^{m}[\widehat{x}]  \bar{Y}_{\ell}^{m}[\widehat{x}'']  \notag \\
	&\hphantom{aaaaa} \times \sum_{\ell',m'}   \frac{( r')^{\ell'}}{(2\ell'+1)!!}    \sum_{s'=0}^{\ell'} \frac{1}{s'! } \frac{(\ell'+s')!}{(\ell'-s')!}  Y_{\ell'}^{m'}[\widehat{x}'']  \bar{Y}_{\ell'}^{m'}[\widehat{x}']  \notag \\
	& \hphantom{aaaaa}  \times \frac{  2^{k+\ell''+1}}{ (-1)^{-k-s-s'+\ell
			+\ell' }} \frac{
		\Gamma [\ell +\ell'+\ell''+2]}{\Gamma [k+s+s'+\ell''+2]}(\eta -\eta')^{-\ell -\ell'-\ell''-2} .
\end{align}

\subsection{$ B_{b,\vec{x}_{o}} $}
The next part is $B_{b,\vec{x}_{o}}$, which is defined as
\begin{align}
	\begin{split}
		B_{b,\vec{x}_{o}} &\equiv \int_{-\infty}^{0}  \dd\eta'' \eta''^{b}  \frac{\delta[\eta-\eta''-|\vec{x}-\vec{x}''|] \delta[\eta''-\eta'-|\vec{x}''-\vec{x}'|]}{(4\pi)^{2}|\vec{x}''-\vec{x}_{o}|},
	\end{split}					
\end{align}
where $b=-1,0$ and $\vec{x}_{o}=\{\vec{x},\vec{x}'\}$.
We can collapse one delta function by carrying out the $\eta''$ integral,
\begin{align}
	\begin{split}
		B_{b,\vec{x}_{o}} &= \left( \eta-|\vec{x}-\vec{x}''| \right)^{b}  \frac{\delta[\eta-\eta'-|\vec{x}''-\vec{x}'|-|\vec{x}-\vec{x}''|]}{(4\pi)^{2}|\vec{x}''-\vec{x}_{o}|} . \\
	\end{split}					
\end{align}
Now, let us use integral representation of delta functions to convert the integrand to frequency space,
\begin{align}
	\begin{split}
		B_{b,\vec{x}_{o}} = \int_{-\infty}^{\infty} \frac{ \dd\omega}{2\pi} \left( \eta-|\vec{x}-\vec{x}''| \right)^{b} e^{-i \omega(\eta-\eta')} \frac{e^{i \omega(|\vec{x}-\vec{x}''|)}e^{i \omega(|\vec{x}''-\vec{x}'|)}}{(4\pi)^{2}|\vec{x}''-\vec{x}_{o}|}.
	\end{split}					
\end{align}
In the next subsection, we will evaluate $B_{b,\vec{x}_{o}}$ for different values of $b$ and $\vec{x}_{o}$.

\subsubsection{$ B_{0,\vec{x}_{o}} $}
For $b=0$, we have
\begin{align}
	\begin{split}
		B_{0,\vec{x}_{o}} &\equiv \int_{-\infty}^{\infty} \frac{\dd\omega}{2\pi}  e^{-i \omega(\eta-\eta')} \frac{e^{i \omega(|\vec{x}-\vec{x}''|)}e^{i \omega(|\vec{x}''-\vec{x}'|)}}{(4\pi)^{2}|\vec{x}''-\vec{x}_{o}|} .
	\end{split}					
\end{align}
We recognize that the integrand is identical to that of $A_{1,2}$. For $\vec{x}_{o}=\vec{x}'$,
\begin{align}
	\begin{split}
		B_{0,\vec{x}'} &= -A_{1,2} ;
		\label{B0a}
	\end{split}					
\end{align}
while for $\vec{x}_{o}=\vec{x}$, we can just swap $r$ and $r'$ from the $A_{1,2}$ result
\begin{align}
		\int_{0}^{\infty} \dd r'' r''^{-\ell''+1} B_{0,\vec{x}}^{\ell' = 0} &=   \frac{  \delta_{\ell',0} }{4\pi} \sum_{\ell,m}  \frac{( r)^{\ell}(\eta -\eta')^{-\ell-\ell''-2}}{(2\ell+1)!!}   \sum_{s=0}^{\ell} \frac{1}{s! } \frac{(\ell+s)!}{(\ell-s)!} Y_{\ell}^{m}[\widehat{x}'']  \bar{Y}_{\ell}^{m}[\widehat{x}]  \notag  \\
		&\hphantom{aaaaa} \times 2^{\ell''-1}   (-1)^{\ell+s} \frac{ (\ell+\ell'')! }{(s+\ell'')!}   \left( \frac{2 r'^{2}}{3}     (\ell+\ell''+1)  +   (\eta -\eta')^{2}  \right) ,\label{B0xl0}
\end{align}
\begin{align}
	&\int_{0}^{\infty}  \dd r'' r''^{-\ell''+1} B_{0,\vec{x}}^{\ell'\geq 1} = \sum_{\ell,m} \sum_{\ell'=1,m'} \sum_{s=0}^{\ell}  \frac{ 2^{\ell''-1} r^{\ell } r'^{\ell'}
		(-1)^{-s+\ell + \ell'} (\eta - \eta')^{-\ell''-\ell -\ell'-2} }{(2 \ell +1) (2 \ell'+1)!!  s! } \frac{ (\ell+s)!}{(\ell-s)!}\label{B0xl1}\\
	&\times \left(\sum_{s'=0}^{\ell'-1}  \frac{(-)^{s'} (\eta-\eta')^{2} \Gamma (s'+\ell' )  \Gamma (\ell''+\ell +\ell')}{(2 \ell' -1)\text{!!} \Gamma (\ell' -s') \Gamma[\ell''+s+s'] s'!}- \sum_{s'=0}^{\ell'+1}  \frac{(-)^{s'} r'^2 \Gamma (s'+\ell' +2) 	\Gamma (\ell''+\ell +\ell'+2)}{(2 \ell' +3)\text{!!}  \Gamma
		[2-s'+\ell' ] \Gamma[\ell''+s+s'] s'!}\right)  . \notag
\end{align}
The first term of eq. \eqref{B0xl0} and second term of eq.\eqref{B0xl1} is suppressed by $(r'/\eta')^{2}$ relative to the 2nd term of eq.~\eqref{B0xl0} and the 1st term of eq.~\eqref{B0xl1}.

\subsubsection{$ B_{-1,\vec{x}_{o}} $}
For $b=-1$, we have
\begin{align}
	\begin{split}
		B_{-1,\vec{x}_{o}} &\equiv \int_{-\infty}^{\infty} \frac{ \dd\omega}{2\pi} \frac{ e^{-i \omega(\eta-\eta')}}{ \eta-|\vec{x}-\vec{x}''| } \frac{e^{i \omega(|\vec{x}-\vec{x}''|)}e^{i \omega(|\vec{x}''-\vec{x}'|)}}{(4\pi)^{2}|\vec{x}''-\vec{x}_{o}|} .
	\end{split}					
\end{align}
At late time $\eta \rightarrow 0$, as long as $|\vec{x}-\vec{x}''| \neq 0$, we can expand $(1-\frac{\eta }{|\vec{x}-\vec{x}''|})^{-1}$ around $\eta=0$. Up to the zeroth order,
\begin{align}
	\begin{split}
		B_{-1,\vec{x}_{o}} 	&\approx \int_{-\infty}^{\infty} \frac{ \dd\omega}{2\pi} \frac{ e^{-i \omega(\eta-\eta')}}{-|\vec{x}-\vec{x}''|  } \frac{e^{i \omega(|\vec{x}-\vec{x}''|)}e^{i \omega(|\vec{x}''-\vec{x}'|)}}{(4\pi)^{2}|\vec{x}''-\vec{x}_{o}|} \left(1 + \mathcal{O}[\frac{\eta }{|\vec{x}-\vec{x}''|} ] \right).
	\end{split}					
\end{align}
For $\vec{x}_{o}=\vec{x}'$, $ B_{-1,\vec{x}'}$ is exactly $A_{0}$ with minus sign difference
\begin{align}
	\begin{split}
		B_{-1,\vec{x}'} &\approx - A_{0} ;
		\label{Bm1a}
	\end{split}					
\end{align}
while, for $\vec{x}_{o}=\vec{x}$, it will be quite different
\begin{align}
	\begin{split}
		B_{-1,\vec{x}} &\approx - \int_{-\infty}^{\infty} \frac{ \dd\omega}{2\pi}  e^{-i \omega(\eta-\eta')} \frac{e^{i \omega(|\vec{x}-\vec{x}''|)}e^{i \omega(|\vec{x}''-\vec{x}'|)}}{(4\pi)^{2}|\vec{x}-\vec{x}''|^{2}} .
	\end{split}					
\end{align}
The use of equations \eqref{DerivativeOfExpROverR} and (\ref{expDivR2}) hands us
\begin{align}
	B_{-1,\vec{x}} &\approx  \int_{-\infty}^{\infty} \frac{ \dd\omega}{2\pi}  e^{-i \omega(\eta-\eta')}  \   \sum_{\ell,m} \frac{(-r\omega)^{\ell }}{(2 \ell+1)!!} \frac{e^{i  \omega r'' }}{r''^{2}} \sum_{s=0}^{\ell} \frac{i^{s+\ell }}{2^{s}s!} \frac{(\ell+s)!}{(\ell-s)!} \sum_{k=0}^{\ell-s}   (-i)^{-k} 	k! (\omega r'')^{-k-s }  C[\ell-s,k] Y_{\ell}^{m}[\widehat{x}]  \bar{Y}_{\ell}^{m}[\widehat{x}''] \notag \\
	&\hphantom{aaaaaaaa} \times  \sum_{\ell',m'}\frac{ r'' r' \omega ^2 }{2 \ell' +1} \left(h_{\ell'
		-1}^{(1)}(r'' \omega ) j_{\ell' -1}(r'
	\omega )-h_{\ell' +1}^{(1)}(r'' \omega ) j_{\ell'
		+1}(r' \omega ) \right)  Y_{\ell'}^{m'}[\widehat{x}'']  \bar{Y}_{\ell'}^{m'}[\widehat{x}']		.
\end{align}
Just as in $A_{1,2}$, we cannot take small argument limit of $j_{-1}[\omega r']$ directly when $\ell'=0$. Because of that, we separate the calculation for $\ell'=0$ and $\ell' \geq 1$. The procedure is the same as previous calculation. By including $r''$, we found
\begin{align}
	\int_{0}^{\infty}  \dd r'' r''^{-\ell''+1} B_{-1,\vec{x}}^{\ell'=0}  &= \frac{\delta_{\ell',0}}{ 12 \pi }  \sum_{\ell,m} \sum_{s=0}^{\ell} \sum_{k=0}^{\ell-s}   k! 2^{k+\ell''} r^{\ell } \frac{(s+\ell )!}{(\ell -s)!} (-1)^{k-s+\ell } \frac{ (\ell	+ \ell'')! (\eta -\eta')^{-\ell -\ell''-3}  C[\ell -s,k]}{ s! (2 \ell +1)!!
		\Gamma [k+s+\ell''+2]} \notag \\
	& \times \left(3 (\eta -\eta')^{2}(1+k+s+\ell'') +2  r'^{2} (\ell +\ell''+1)(\ell +\ell''+2) \right) Y_{\ell}^{m}[\widehat{x}]  \bar{Y}_{\ell}^{m}[\widehat{x}'']	.
\end{align}
\begin{align}
	&\int_{0}^{\infty}  \dd r'' r''^{-\ell''+1} B_{-1,\vec{x}}^{\ell' \geq 1} \notag \\
	&\hphantom{aaaa} =\sum_{\ell,m} \sum_{s=0}^{\ell} \sum_{k=0}^{\ell-s} \sum_{\ell'=1,m'} \left( \sum_{s'=1}^{\ell'-1}  (\eta -\eta')^2 (\ell'-s') \frac{(\ell'-s'+1)}{(2 \ell'-1)!!}\Gamma [\ell'+s'] \Gamma[\ell+\ell'+\ell''+1] \right. \notag \\
	&\hphantom{aaaaaaaaaaaaaaaaaaa}\left. - \sum_{s'=1}^{\ell'+1}  r'^2  \Gamma [\ell'+s'+2]
	\frac{\Gamma[\ell+\ell'+\ell''+3]}{(2 \ell'+3)!!}  \right) \notag \\
	&\hphantom{aaaaaa} \times \frac{(-i)^{3 \ell} 2^{k+\ell''} \Gamma [k+1] (i r)^{\ell} r'^{\ell'} \Gamma[\ell+s+1]
		(-1)^{-k+\ell'-s-s'}  (\eta - \eta')^{-\ell-\ell'-\ell''-3}
		C[\ell-s,k] }{(2 \ell'+1) s!
		s'! (2 \ell+1)!!   (\ell-s)! \Gamma (\ell'-s'+2)
		\Gamma[k+\ell'+s+s'+1]}  .
\end{align}

\subsection{$ C_{-1} $}
The last one is $C_{-1}$:
\begin{align}
	\begin{split}
		C_{-1} &= \int_{-\infty}^{0}  \dd\eta'' \frac{1}{(4\pi)^{2}\eta''}  \delta[\eta-\eta''-|\vec{x}-\vec{x}''|] \delta[\eta''-\eta'-|\vec{x}''-\vec{x}'|] .
	\end{split}					
\end{align}
We may evaluate the $\eta''$ integral by collapsing one of the delta functions, and use the delta function's integral representation to convert the integrand to frequency space,
\begin{align}
	\begin{split}
		C_{-1} &=\frac{1}{(4\pi)^{2}} \int_{-\infty}^{\infty} \frac{ \dd\omega}{2\pi} \frac{ e^{-i \omega(\eta-\eta')}}{ \eta-|\vec{x}-\vec{x}''| } e^{i \omega(|\vec{x}-\vec{x}''|)}e^{i \omega(|\vec{x}''-\vec{x}'|)} .
	\end{split}					
\end{align}
At late time $\eta \rightarrow 0$, as long as $|\vec{x}-\vec{x}''| \neq 0$, we can expand $(1-\frac{\eta }{|\vec{x}-\vec{x}''|})^{-1}$ around $\eta=0$. Up to zeroth order,
\begin{align}
	\begin{split}
		C_{-1} 	&\approx -\int_{-\infty}^{\infty} \frac{ \dd\omega}{2\pi} \frac{ e^{-i \omega(\eta-\eta')}}{(4\pi)^{2}|\vec{x}-\vec{x}''|  } e^{i \omega(|\vec{x}-\vec{x}''|)}e^{i \omega(|\vec{x}''-\vec{x}'|)} \left(1 + \mathcal{O}[\frac{\eta }{|\vec{x}-\vec{x}''|} ] \right) = -B_{0,\vec{x}} .
		\label{Cm1a}
	\end{split}					
\end{align}

\end{document}